\documentclass[journal=jctcce,manuscript=article,
]{achemso}
\SectionNumbersOn

\usepackage{amsmath}
\usepackage{amsfonts}
\usepackage{braket}
\usepackage{graphicx}
\usepackage{subcaption}
\usepackage{booktabs}
\usepackage[american]{babel}
\usepackage{multirow}
\usepackage{mathtools}

\usepackage{listings}
\usepackage{xcolor}

\definecolor{codegreen}{rgb}{0,0.6,0}
\definecolor{codegray}{rgb}{0.5,0.5,0.5}
\definecolor{codepurple}{rgb}{0.58,0,0.82}
\definecolor{backcolour}{rgb}{0.95,0.95,0.92}

\lstdefinestyle{mystyle}{
  backgroundcolor=\color{backcolour},
  commentstyle=\color{codegreen},
  keywordstyle=\color{magenta},
  numberstyle=\tiny\color{codegray},
  stringstyle=\color{codepurple},
  basicstyle=\ttfamily\footnotesize,
  breakatwhitespace=false,
  breaklines=true,
  captionpos=b,
  keepspaces=true,
  numbers=left,
  numbersep=5pt,
  showspaces=false,
  showstringspaces=false,
  showtabs=false,
  tabsize=2
}

\usepackage[colorlinks,linkcolor=black,urlcolor=blue, citecolor=black]{hyperref} 

\newcommand{\KOFO}{Max-Planck-Institut f{\"u}r Kohlenforschung, Kaiser-Wilhelm-Platz 1, 45470 M{\"u}lheim an der Ruhr, Germany}
\newcommand{\ETH}{ETH Z{\"u}rich, Department of Chemistry and Applied Biosciences, Vladimir-Prelog-Weg 2,
8093 Z{\"u}rich, Switzerland}

\title{
  Efficient Implementation of the Spin-Free Renormalized Internally-Contracted Multireference Coupled Cluster Theory
}
\author{Kalman Szenes}
\affiliation{\ETH}
\author{Riya Kayal}
\affiliation{\KOFO}
\author{Kantharuban Sivalingam}
\affiliation{\KOFO}
\author{Robin Feldmann}
\affiliation{\ETH}
\author{Frank Neese}
\email{neese@kofo.mpg.de}
\affiliation{\KOFO}
\author{Markus Reiher}
\email{mreiher@ethz.ch}
\affiliation{\ETH}

\date{05/11/2025}

\begin{document}

\begin{abstract}
  In this paper, an efficient implementation of the renormalized internally-contracted multreference coupled cluster with singles and doubles (RIC-MRCCSD) into the ORCA quantum chemistry program suite is reported.
  To this end, Evangelista's \texttt{Wick\&d} equation generator was combined with ORCA's native \texttt{AGE} code generator in order to implement the many-body residuals required for the RIC-MRCCSD method.
  Substantial efficiency gains are realized by deriving a spin-free formulation instead of the previously reported spin-orbital version developed by some of us.
  Since \texttt{AGE} produces parallelized code, the resulting implementation can directly be run in parallel with substantial speedups when executed on multiple cores.
  In terms of runtime, the cost of RIC-MRCCSD is shown to be between single-reference RHF-CCSD and UHF-CCSD, even when active space spaces as large as CAS(14,14) are considered.
  This achievement is largely due to the fact that no reduced density matrices (RDM) or cumulants higher than three-body enter the formalism.
  The scalability of the method to large systems is furthermore demonstrated by computing the ground-state of a vitamin B\textsubscript{12} model comprised of an active space of CAS(12, 12) and 809 orbitals. 
  In terms of accuracy, RIC-MRCCSD is carefully compared to second- and approximate fourth-order $n$-electron valence state perturbation theories (NEVPT2, NEVPT4(SD)), to the multireference zeroth-order coupled-electron pair approximation (CEPA(0)), as well as to the IC-MRCCSD from K\"ohn.
  In contrast to RIC-MRCCSD, the IC-MRCCSD equations are entirely derived by \texttt{AGE} using the conventional projection-based approach, which, however, leads to much higher algorithmic complexity than the former as well as the necessity to calculate up to the five-body RDMs.
  Remaining challenges such as the variation of the results with the flow, a free parameter that enters the RIC-MRCCSD theory, are discussed.

\end{abstract}

\section{Introduction}

The electronic structure of closed-shell molecules can be accurately described by coupled cluster theory~\cite{shavittManyBodyMethodsChemistry2009}, which relies on the mean-field Hartree-Fock determinant as a zeroth-order approximation.
However, for many chemically relevant systems, such as transition-metal complexes and biradicals, the mean-field solution does not dominate the full wavefunction and, consequently, methods built upon it suffer in terms of reliability and accuracy.
Therefore, such systems are typically described with active orbital space methods, which partition the orbitals into subspaces, of which one is usually treated exactly while the remaining orbitals are neglected.

If the orbitals are chosen carefully, such that strongly correlated orbitals are included in the subspace, this scheme can provide a qualitative description of the full electronic structure.
However, to obtain accurate properties, such active space methods need to be complemented by multireference schemes, which attempt to recover the electron correlation from the full set of orbitals, including the neglected orbitals.
The workhorse of these methods in chemistry is second-order perturbation theory, for which complete active space (CASPT2)~\cite{Andersson1990Jul} and $n$-electron valence state perturbation theory (NEVPT2)~\cite{Angeli2001Jun} are the two prime examples.
These methods are currently considered state of the art despite providing only low-order perturbative corrections.

By contrast, a variety of multireference coupled cluster methods exists to address complicated open-shell systems, ranging from an uncontracted ansatz (Jezorski-Monkhorst)~\cite{jeziorskiCoupledclusterMethodMultideterminantal1981} to fully internally-contracted ones.\cite{hanauerPilotApplicationsInternally2011,evangelista_sequential_2012,yanai_canonical_2006} These methods are computationally demanding and thus limited to niche applications.~\cite{lyakhMultireferenceNatureChemistry2012,evangelistaPerspectiveMultireferenceCoupled2018}
Among them, the internally-contracted MRCC (IC-MRCC) developed by K\"ohn and coworkers presents a very general formalism.~\cite{hanauerPilotApplicationsInternally2011,hanauerPerturbativeTreatmentTriple2012,evangelistaSequentialTransformationApproach2012,Samanta2014Apr,aotoInternallyContractedMultireference2016,blackEfficientImplementationInternally2023}
However, the ansatz leads to reduced density matrices of up to fifth order, which severely limits its application to small active spaces.
Internally-contracted coupled-electron pair approximation (CEPA) approaches are more affordable and similar in quality, although some variants may lack formal properties such as size-consistency or orbital invariance.~\cite{blackLinearQuadraticInternally2019}

Higher order reduced matrices can be avoided by exploiting many-body residuals, as in the partially internally-contracted MRCC approach of Datta and Nooijen.~\cite{dattaStatespecificPartiallyInternally2011} However, this formalism can suffer from convergence issues, as reported by Lechner and coworkers.~\cite{lechnerPerturbativeApproachMultireference2021}
Driven similarity renormalization group (DSRG) approaches, developed in the group of Evangelista, mitigate these issues by regularizing diverging amplitudes.~\cite{Evangelista2014Aug,liMultireferenceDrivenSimilarity2015,liNumericallyRobustMultireference2016}
In Ref.~\citenum{feldmannRenormalizedInternallyContracted2024}, some of us introduced the renormalized internally-contracted multireference coupled cluster (RIC-MRCC) theory, a novel multireference approach that adapts the DSRG scheme to non-unitary similarity transformations, thereby resembling conventional MRCC methods.~\cite{Cizek1969Jan,mukherjee1975correlation,Banerjee1981Feb,Banerjee1982May,Laidig1984Feb,Meller1996Mar,evangelistaOrbitalinvariantInternallyContracted2011,hanauerPilotApplicationsInternally2011}
Its defining characteristics can be summarized as follows:
\begin{enumerate}
  \item RIC-MRCC relies on the internally-contracted ansatz, where a single cluster operator is applied on the entire CAS reference wavefunction.
  \item It relies on many-body residuals~\cite{dattaStatespecificPartiallyInternally2011} based on the generalized normal-ordering of Mukherjee and Kutzelnigg~\cite{Mukherjee1997Aug,kutzelniggNormalOrderExtended1997}, which correspond to the matrix elements of the effective Hamiltonian.
    This yields a simpler set of residual equations compared to the conventional ones obtained from the projected residuals.
    These residuals are devoid of linear dependencies, which are inherent to the internally-contracted ansatz~\cite{Werner1988Nov,Andersson1990Jul} and plague MRCC schemes in particular.~\cite{kohnImprovedSimplifiedOrthogonalisation2020,blackEfficientImplementationInternally2023,leeSpinfreeGeneralisedNormal2025}
  \item To simplify the working equations, a large number of contractions involving amplitudes with multiple active indices have been neglected.
    As a consequence, RIC-MRCC only relies on up to three-body reduced density matrices and cumulants, making it amenable to large active space calculations.
  \item The update equation for the amplitudes is augmented by a regularization factor, which attempts to remove numerical instabilities.
    This factor, however, introduces a free parameter which is examined in this study.
\end{enumerate}

In this paper, we present an implementation of the RIC-MRCC method with single and double excitations in the ORCA~\cite{Neese2020Jun} quantum chemistry package.
By contrast to the initial publication\cite{feldmannRenormalizedInternallyContracted2024}, which was expressed in a spin-orbital basis, we reformulate all equation in spin-free form.
To achieve this, the many-body residual equations obtained from the \texttt{Wick\&d}~\cite{evangelistaAutomaticDerivationManybody2022} program have been translated to ORCA's internal code generator \texttt{AGE}'s~\cite{lechnerCodeGenerationORCA2024} format, which subsequently carries out the spin adaptation of the equations.
We evaluate the performance of our method both in terms of accuracy and efficiency compared to other widely used single- and multireference schemes for a range of closed- and open-shell systems.

This manuscript is organized as follows: Section~\ref{sec:theory} first reviews the RIC-MRCC method and then presents the principles behind the spin-free formulation.
Subsequently, Section~\ref{sec:implementation} provides implementation details of the method in ORCA, including the translation layer that was developed between the two code generators \texttt{Wick\&d} and \texttt{AGE}.
In Section~\ref{sec:results}, numerical results are presented, including computational timings compared to state-of-the-art single- and multireference methods, accuracy on a benchmark of transition-metal ions and a large-scale calculation on the vitamin B\textsubscript{12} model.
In addition, the effect of the free parameter in the regularization factor on the stability and accuracy of the method is investigated.
Finally, Section~\ref{sec:conclusion} concludes the paper with a summary of the main findings and an outlook for future work.

\section{Theory}\label{sec:theory}

\subsection{Generalized Normal-Ordering}

The generalized normal-ordering (GNO) formalism developed by Mukherjee and Kutzelnigg~\cite{Mukherjee1997Aug,kutzelniggNormalOrderExtended1997} extends the concept of normal-ordering to general multi-determinantal vacuum states $\ket{\Psi_{0}}$.
Although this framework is applicable to arbitrary wavefunctions, in this study, $\ket{\Psi_{0}}$ is assumed to originate from a complete active space (CAS) procedure~\cite{roos1980complete,roos1980complete2}
\begin{equation}
  \ket{\Psi_{0}} = \sum_{t=1}^{N_{\text{CAS}}} c_{t} \ket{\phi_{t}},
\end{equation}
where $t$ runs over the $N_{\text{CAS}}$ many-body expansion functions, such as configuration state functions (CSF) or Slater determinants, $\ket{\phi_{t}}$ in the active space.
This manuscript relies on conventions for the orbital indices which are summarized in Table~\ref{tab:space-convention}.
\begin{table}[htb]
  \centering
  \begin{tabular}{lccc}
    \toprule
    \textbf{Space} & \textbf{Symbol} & \textbf{Indices} & \textbf{Definition} \\
    \midrule
    Internal      & $\mathbb{C}$ & $i, j$                & occupied \\
    Active        & $\mathbb{A}$ & $t, u, v, x, y, z$    & active \\
    Virtual       & $\mathbb{V}$ & $a, b$                & unoccupied \\
    Hole          & $\mathbb{H}$ & $k, l, m, n$          & $\mathbb{H} = \mathbb{C} \cup \mathbb{A}$ \\
    Particle      & $\mathbb{P}$ & $c, d, e, f$          & $\mathbb{P} = \mathbb{A} \cup \mathbb{V}$ \\
    General       & $\mathbb{G}$ & $p, q, r, s$          & $\mathbb{G} = \mathbb{H} \cup \mathbb{V}$ \\
    \bottomrule
  \end{tabular}
  \caption{Orbital space decomposition and corresponding index conventions}
  \label{tab:space-convention}
\end{table}

Adopting the terse notation for second-quantized operators in spin-orbital basis
\begin{equation}
  \hat{a}^{pq \ldots}_{rs \ldots} = \hat{a}^{\dagger}_{p} \hat{a}^{\dagger}_{q} \ldots \hat{a}_{s} \hat{a}_{r},
\end{equation}
the contractions arising from the generalized Wick's theorem~\cite{Mukherjee1997Aug,kutzelniggNormalOrderExtended1997,Kong2010Jun} yield the one-particle reduced density matrix (RDM) $\gamma^{t}_{u} = \braket{ \Psi_{0}  | \hat{a}^{t}_{u} | \Psi_{0}  }$ and the one-hole density matrix $\eta_{u}^{t} = \delta_{u}^{t} - \gamma_{u}^{t}$, along with \emph{multi-legged} contractions producing $n$-body density cumulants~\cite{kutzelniggCumulantExpansionReduced1999}, which are composed of antisymmetrized products of $n$- and lower-body RDMs.
For the two-body case, it takes the form
\begin{equation}\label{eq:cumulant}
  \lambda^{tu}_{vx} = \gamma^{tu}_{vx} - \gamma^{t}_{v}\gamma^{u}_{x} + \gamma^{t}_{x}\gamma^{u}_{u}.
\end{equation}
A concise summary of the contractions rules arising from the GNO formalism can be found in Ref.~\citenum{evangelistaAutomaticDerivationManybody2022}, while the original publications by Mukherjee and Kutzelnigg~\cite{Mukherjee1997Aug,kutzelniggNormalOrderExtended1997} provide a thorough discussion.

On this basis, the Born--Oppenheimer electronic Hamiltonian can be expressed in normal-ordered form with respect to $\ket{\Psi_{0}}$ as
\begin{equation}
  \hat{H} = E_{0} + \sum_{pq}^{\mathbb{G}} f^{p}_{q}\{\hat{a}^{p}_{q}\} + \frac{1}{4}\sum_{pqrs}^{\mathbb{G}} v^{pq}_{rs}\{\hat{a}^{pq}_{rs}\},
\end{equation}
where $\{\cdot\}$ denotes normal-ordered operators.
The scalar term comprises the reference energy $E_{0} = \braket{ \Psi_{0} | \hat{H} | \Psi_{0} }$, and the one-electron part is given by the generalized Fock matrix~\cite{helgakerMolecularElectronicstructureTheory2000}
\begin{equation}
  f^{p}_{q} = h^{p}_{q} + \sum_{m}^{\mathbb{C}} v^{pm}_{qm} + \sum_{tu}^{\mathbb{A}} v^{pt}_{qu} \gamma^{t}_{u}.
\end{equation}
Here, $h^{p}_{q} = \braket{ \phi_{p} | \hat{h} | \phi_{q} }$ and $v^{pq}_{rs} = \braket{ pq || rs }$ correspond to the standard one-electron and anstizymmetrized two-electron integrals (in $\braket{ 12 | 12 }$ physics notation), respectively.

\subsection{Renormalized Internally-Contracted Multireference Coupled Cluster Theory}
\label{sec:ricmrcc}

The renormalized internally-contracted multireference coupled cluster theory~\cite{feldmannRenormalizedInternallyContracted2024} (RIC-MRCC) introduced by some of the authors of this work is a natural extension of Evangelista's driven similarity renormalization group (DSRG) theory~\cite{Evangelista2014Aug,liMultireferenceDrivenSimilarity2015,liNumericallyRobustMultireference2016} with the key difference of relying on a non-unitary similarity transformations, thereby resembling internally-contracted multireference coupled cluster (IC-MRCC)~\cite{Cizek1969Jan,mukherjee1975correlation,Banerjee1981Feb,Banerjee1982May,Laidig1984Feb,Meller1996Mar,evangelistaOrbitalinvariantInternallyContracted2011,hanauerPilotApplicationsInternally2011} schemes.
Instead of repeating the derivation of this theory starting from DSRG---as presented in the original publication~\cite{feldmannRenormalizedInternallyContracted2024}---this work relates the working equations directly to the ones from IC-MRCC.

In IC-MRCC, the wavefunction ansatz is defined by applying the cluster operator $e^{\hat{T}}$ to the reference CAS solution
\begin{equation}
  \ket{\Psi_{\text{IC-MRCC}}} = e^{\hat{T}} \ket{\Psi_{0}}.
\end{equation}
The cluster operator is decomposed in terms of $n$-body excitation operators, typically truncated at a chosen excitation rank.
In the initial work on RIC-MRCC~\cite{feldmannRenormalizedInternallyContracted2024}, a perturbative triples correction was introduced---based on the work of Hanauer and K\"ohn~\cite{hanauerPerturbativeTreatmentTriple2012}---on top of the iterative singles and doubles solution.
This extension, however, lies beyond the scope of the present study and will be addressed elsewhere.
Hence, in this work, $\hat{T}$ will be restricted to single and double excitations from the hole ($\mathbb{H}$) to the particle ($\mathbb{P}$) space
\begin{equation}
  \hat{T} = \hat{T}_{1} + \hat{T}_{2} = \sum_{k}^{\mathbb{H}}\sum_{c}^{\mathbb{P}}  t^{c}_{k}\{\hat{a}_{k}^{c}\} + \frac{1}{4} \sum_{kl}^{\mathbb{H}}\sum_{cd}^{\mathbb{P}}  t^{cd}_{kl} \{\hat{a}^{cd}_{kl}\}.
\end{equation}
Excitations involving only active indices are omitted from the cluster operator.
Their influence is confined to modifying the CAS expansion coefficients and can therefore be regarded as a reference relaxation effect.
In IC-MRCC schemes, this phenomenon is typically accounted for by solving the CAS problem for the effective Hamiltonian $\overline{H}$~\cite{evangelistaOrbitalinvariantInternallyContracted2011,hanauerPilotApplicationsInternally2011}, an approach that we intend to incorporate in future work.
The effective Hamiltonian is obtained by similarity transformation of the Born--Oppenheimer Hamiltonian by the cluster operator
\begin{equation}
  \overline{H} = e^{-\hat{T}} \hat{H} e^{\hat{T}}.
\end{equation}
In conventional CC theory, the energy $E_{\text{CC}}$ and the singles $r^{c}_{k}$ and doubles $r^{cd}_{kl}$ residual equations are derived by projecting the Schr\"odinger equation
\begin{equation}
  \overline{H} \ket{\Psi_{0}} = E_{\text{CC}} \ket{\Psi_{0}}
\end{equation}
onto the reference and internally-contracted excited configurations, respectively
\begin{align}
  E_{\text{CC}} &\coloneqq \braket{ \Psi_{0} | \overline{H} | \Psi_{0} }, \\
  r^{c}_{k} &\coloneqq \braket{ \Psi_{0} | \hat{a}^{k}_{c} \overline{H} | \Psi_{0} } \overset{!}{=} 0, \\
  r^{cd}_{kl} &\coloneqq \braket{ \Psi_{0} | \hat{a}^{kl}_{cd} \overline{H} | \Psi_{0} } \overset{!}{=} 0.
\end{align}
Note, however, that the excited configurations $\hat{a}^{c}_{k} \ket{\Psi_{0}}$ and $\hat{a}^{cd}_{kl} \ket{\Psi_{0}}$ are, in general, linearly dependent, which often results in numerical instabilities during the iterative optimization procedure.
This is an issue inherent to the IC ansatz and is typically addressed in MR configuration interaction and perturbation theories by defining a linearly independent set of excitation operators through canonical orthogonalization.~\cite{Werner1988Nov,Andersson1990Jul}
Careful treatment of this redundancy---particularly for the single particle excitations---is necessary in IC-MRCC to ensure orbital-invariance~\cite{evangelistaOrbitalinvariantInternallyContracted2011} and size-extensivity~\cite{hanauerPilotApplicationsInternally2011,hanauerCommunicationRestoringFull2012} of the method, aspects that have been the subject of thorough study in the literature.~\cite{kohnImprovedSimplifiedOrthogonalisation2020,blackEfficientImplementationInternally2023,leeSpinfreeGeneralisedNormal2025}
In addition, these schemes typically employ thresholds for discarding linearly dependent terms, which may introduce discontinuities in the energies along a potential energy surface.~\cite{blackLinearQuadraticInternally2019}

An alternative approach for defining the residual equations relies on expanding $\overline{H}$ in terms of contributions grouped by $n$-body operators
\begin{equation}
  \overline{H} = \overline{H}_{0} + \sum_{pq}^{\mathbb{G}}\overline{H}_{q}^{p}\{\hat{a}^{p}_{q}\} + \frac{1}{4} \sum_{pqrs}^{\mathbb{G}}\overline{H}_{rs}^{pq}\{\hat{a}^{pq}_{rs}\} + \ldots
\end{equation}
Here, the energy is provided by the zeroth-order term---as all other terms contain normal-ordered operators which vanish when evaluating reference expectation values---while the singles and doubles residuals correspond to the effective one- and two-body components
\begin{align}
  E_{\text{CC}} &\coloneqq \overline{H}_{0}, \\
  r^{c}_{k} &\coloneqq \overline{H}^{c}_{k} \overset{!}{=} 0, \\
  r^{cd}_{kl} &\coloneqq \overline{H}^{cd}_{kl} \overset{!}{=} 0,
\end{align}
yielding the \emph{many-body residuals}, a term coined in Ref.~\citenum{dattaStatespecificPartiallyInternally2011}.
The RIC-MRCC method relies on this form of the residuals and the working equations are derived using the GNO formalism.
For single-determinantal reference wavefunctions, the projected and many-body residual formulations lead to identical equations.~\cite{shavittManyBodyMethodsChemistry2009}
This equivalence does not hold, however, for multireference wavefunctions, where the many-body residual equations form a simpler set of equations than the projected ones.~\cite{dattaStatespecificPartiallyInternally2011,lechnerInternallyContractedMultireference}
An additional advantage of this formalism is that the resulting residuals are devoid of any redundancy even for linearly dependent amplitudes.~\cite{dattaStatespecificPartiallyInternally2011}

To obtain the working equations, $\overline{H}$ is expanded according to the Baker--Campbell--Hausdorff (BCH)
formula as
\begin{equation}
  \overline{H} = \hat{H} + [\hat{H}, \hat{T}] + \frac{1}{2} [[\hat{H}, \hat{T}], \hat{T}] + \ldots,
\end{equation}
which we truncate at the two-fold commutator, an approximation that has been shown to have a negligible effect on the accuracy of IC-MRCC methods.~\cite{evangelistaOrbitalinvariantInternallyContracted2011,hanauerPilotApplicationsInternally2011}
Even with this truncation, however, evaluating all the resulting contractions becomes impractical for anything beyond small, few-electron systems.
Therefore, RIC-MRCC employs a set of additional approximations that neglect costly contractions involving amplitudes with active orbital indices from the two-fold commutator $[[\hat{H}, \hat{T}], \hat{T}]$.
As in Ref~\citenum{blackLinearQuadraticInternally2019}, distinct approximation are applied for the energy and residual equations:
\begin{itemize}
  \item Energy contribution $(E_{\text{CC}})$: contractions involving multiple amplitudes with three active indices are omitted.
  \item Residual contribution $(r^{c}_{k}$, $r_{kl}^{cd})$: contractions involving multiple amplitudes with active indices, as well as all those containing the two-body cumulant, are neglected.
\end{itemize}
Although these approximations have primarily been chosen to decrease the computational cost of the method, physical motivations for them can be found in the initial publication on RIC-MRCC.~\cite{feldmannRenormalizedInternallyContracted2024}
An important consequence of these simplifications is that, unlike the untruncated equations that depend on up to four-body cumulants, RIC-MRCC requires only up to three-body cumulants, as all contractions involving the four-body cumulant are omitted through the scheme.
This reduction is especially beneficial for systems with large active spaces, where evaluating higher-order RDMs---and their associated cumulants---constitutes the primary bottleneck both in terms of computational cost and memory usage.

The resulting coupled cluster equations are solved using a direct inversion of the iterative subspace~\cite{Pulay1980Jul,Pulay1982Dec,Scuseria1986Oct} accelerated quasi-Newton iterative procedure
\begin{equation}\label{eq:cc-update}
  t_{\nu} \leftarrow t_{\nu} + \frac{r_{\nu}}{\Delta_{\nu}},
\end{equation}
where the compound index $\nu$ encompasses particle (upper) and hole (lower) indices and the preconditioner is given by the generalized M{\o}ller--Plesset denominators $\Delta_{\nu} \coloneqq \Delta^{kl \ldots}_{cd \ldots} \coloneqq \epsilon_{k} + \epsilon_{l} + \ldots - \epsilon_{c} - \epsilon_{d}$, corresponding to diagonal elements of the generalized Fock matrix.

Clearly, this iterative procedure can suffer from numerical instabilities~\cite{dattaStatespecificPartiallyInternally2011,dattaMultireferenceEquationofmotionCoupled2012} in the case of small or vanishing denominators that cause the second term in Eq.~\eqref{eq:cc-update} to diverge.
To mitigate this problem in RIC-MRCC, the amplitude update rule is augmented by a renormalization factor
\begin{equation}\label{eq:ric-update}
  t_{\nu} \leftarrow (t_{\nu}\Delta_{\nu} + r_{\nu})\frac{1 - e^{-s\Delta_{\nu}^2}}{\Delta_{\nu}}.
\end{equation}
This modification ensures that even for problematic vanishing denominators the updated amplitudes remain bounded.
Originating from DSRG theory, such renormalization factors have also been incorporated into single- and multireference perturbation theories such as regularized MP2~\cite{lee2018regularized} and CASPT2~\cite{battaglia2022regularized}.
In the context of CASPT2, it serves precisely the same role as the well-known real~\cite{Roos1995Oct} and imaginary~\cite{forsberg1997multiconfiguration} shift parameters for mitigating intruder states.

A notable feature of our theory is that, stemming from DSRG, the regularization factor requires semi-canonical orbitals---those that diagonalize the generalized Fock operator in the internal, active, and virtual spaces separately.
The method can, in principle, be made orbital-invariant at the expense of spoiling the simple structure of the regularization factor in Eq.~\eqref{eq:ric-update}.~\cite{liMultireferenceTheoriesElectron2019}
As a state-specific approach, this orbital canonicalization must therefore be performed individually for each electronic state of interest.

\subsection{Deriving Spin-Free Equations}

Our earlier work~\cite{feldmannRenormalizedInternallyContracted2024} implemented the working equations for the RIC-MRCC method in a spin-orbital basis.
In this work, these equations are reformulated in spin-free form, following the procedure outlined in Ref.~\citenum{dattaStatespecificPartiallyInternally2011} for the many-body residual formulation of IC-MRCC.
The core idea is to identify relations between the coefficients of different spin components of tensors, allowing the reduction of spin-orbital quantities to a single representative set of spin indices from which all other spin sectors can be recovered.
In the past, this principle has been applied to derive certain spin-adapted single-reference CC schemes.~\cite{matthewsRevisitationNonorthogonalSpin2013,matthewsNonorthogonalSpinadaptationCoupled2015,scuseriaClosedshellCoupledCluster1987,nooijen1996general}
More recently, this approach has found applications in the context of multireference methods where it has been used to derive both many-body~\cite{dattaStatespecificPartiallyInternally2011} and projected~\cite{liSpinfreeFormulationMultireference2021} spin-free variants of IC-MRCC as well as DSRG equations~\cite{liSpinfreeFormulationMultireference2021}.

The next two subsections establish fundamental properties of general antisymmetric singlet tensors, which enables us to identify a minimal set of non-redundant spin-orbital components.
These coefficients are subsequently expressed in terms of spin-free quantities, allowing the removal of all spin labels from the tensors present in the contractions.
Then, we demonstrate that all tensors involved in the contractions can be indeed considered as singlets, thereby justifying their replacement by their spin-free counterparts.
Our derivation closely follows the treatment given in the appendix of Ref.~\citenum{shamasundarCumulantDecompositionReduced2009}.

\subsubsection{Singlet Constraining Conditions for Antisymmetric Tensors}\label{subsec:singlet-constrain}

Consider a general $n$-body operator
\begin{equation}\label{eq:asym-op}
  \hat{O} = \sum_{p q  \ldots} \sum_{\substack{\sigma_{1} \sigma_{2} \ldots \\ \tau_{1} \tau_{2} \ldots}}^{\{\alpha, \beta\}} o^{p_{\sigma_{1}} q_{\sigma_{2}} \ldots}_{r_{\tau_{1}} s_{\tau_{2} \ldots}}  \hat{a}^{p_{\sigma_{1}} q_{\sigma_{2}} \ldots}_{r_{\tau_{1}} s_{\tau_{2}} \ldots},
\end{equation}
where $\{p, q, \ldots\}$ identify spatial orbitals and $\{\sigma_{1}, \sigma_{2},\ldots\}, \{\tau_{1}, \tau_2 \ldots \}$ label $\alpha$ and $\beta$ spin components.
We adopt the convention that lowercase indices refer to spin-orbitals, and uppercase indices to spatial orbitals.
If a spin label is omitted from a spin-orbital index, the index is assumed to correspond to the $\alpha$ component while an overbar denotes the $\beta$ component.
Additionally, note that, within this subsection, the indices $t$ and $u$ refer to general indices instead of active ones.

The operator in Eq.~\eqref{eq:asym-op} is considered \emph{antisymmetric} if its coefficients $o_{r_{\tau_{1}} s_{\tau_2 } \ldots}^{p_{\sigma_1 } q_{\sigma_2} \ldots}$ are antisymmetric under permutations of either upper or lower indices
\begin{equation}\label{eq:antisym-op}
  o_{r_{\tau_1 } s_{\tau_2 } \ldots}^{p_{\sigma_1 } q_{\sigma_2} \ldots} = - o_{r_{\tau_1 } s_{\tau_2 } \ldots}^{ q_{\sigma_2} p_{\sigma_1 } \ldots} = - o_{s_{\tau_2 } r_{\tau_1 }  \ldots}^{ p_{\sigma_1 } q_{\sigma_2}  \ldots} = o_{s_{\tau_2 } r_{\tau_1 }  \ldots}^{  q_{\sigma_2} p_{\sigma_1 }  \ldots}
\end{equation}
Additionally, for the tensor to constitute a \emph{singlet}, it must commute with the three standard spin angular momentum operators $[\hat{S}_{+}, \hat{O}] = [\hat{S}_{-}, \hat{O}] = [\hat{S}_{z}, \hat{O}] = 0$.
By explicitly evaluating these commutators, one can derive \emph{singlet constraints}~\cite{shamasundarCumulantDecompositionReduced2009}, which force certain coefficients to vanish and relate coefficients of different spin sectors to each other.
The key results of this procedure are
\begin{itemize}
  \item $[\hat{S}_{z}, \hat{O}] = 0$ implies that coefficients that do not conserve the $M_{S}$ quantum number---coefficients where the number of $\alpha$ and $\beta$ indices in the upper and lower sets of indices differ---must vanish.
  \item Evaluating $[\hat{S}_{-}, \hat{O}]$ yields the following relations:
    \begin{enumerate}
      \item Coefficients where pairs of lower and upper $\alpha$/$\beta$ indices are exchanged are equivalent:
        \begin{equation}\label{eq:one-body-beta2alpha}
          \begin{aligned}
            &\text{1-body:} & o^{p}_{q} &= o^{\overline{p}}_{\overline{q}}, \\
            &\text{2-body:} & o^{\overline{p} q}_{\overline{r} s} &= o^{p \overline{q}}_{r \overline{s}},
            \quad o^{pq}_{rs} = o^{\overline{pq}}_{\overline{rs}}, \\
            &\text{3-body:} & o^{pqr}_{stu} &= o^{\overline{pqr}}_{\overline{stu}},
            \quad o^{\overline{pq} r}_{\overline{st} u} = o^{pq\overline{r}}_{st\overline{u}}.
          \end{aligned}
        \end{equation}
      \item Coefficients containing only $\alpha$ spin indices can be expressed in terms of those containing a single $\beta$ pair:
        \begin{equation}\label{eq:all-alphas}
          \begin{aligned}
            & \text{2-body:} & o^{pq}_{rs}   & = o^{p \overline{q}}_{r \overline{s}} - o^{q \overline{p}}_{s \overline{r}}, \\
            & \text{3-body:} & o^{pqr}_{stu} & = o^{p q \overline{r}}_{s t \overline{u}} - o^{p r \overline{q}}_{s t \overline{u}} - o^{r q \overline{p}}_{s t \overline{u}}.
          \end{aligned}
        \end{equation}
        Owing to the relation between the first and last coefficients in Eq.~\eqref{eq:antisym-op}, these properties hold when permuting the upper indices, as shown in Eq.~\eqref{eq:all-alphas}, as well as when permuting the lower indices.
    \end{enumerate}
\end{itemize}
These relations demonstrate that for one-, two-, and three-body operators, there exists only a single unique spin pattern---$o^{p}_{q}$, $o^{p \overline{q}}_{r \overline{s}}$, and $o^{pq \overline{r}}_{st \overline{u}}$, respectively---from which all other spin components can be obtained.
Taking this into account dramatically reduces the number of non-redundant equations present in the spin-orbital form of the contractions.
Note that these relations are entirely general, with no assumptions made on the nature of the tensor beyond its antisymmetry and singlet property.

Having identified a list of non-redundant spin-orbital quantities, the goal now is to express them in terms of spin-free quantities.
The corresponding spin-free tensor coefficients $O_{R S \ldots}^{P Q \ldots}$ can be obtained by integrating out the spin degrees of freedom~\cite{Kutzelnigg1982Sep}
\begin{align}\label{eq:spinfree-spinsum}
  \hat{O} &= \sum_{PQRS \ldots} O^{P Q \ldots}_{RS \ldots}  \hat{E}_{P Q \ldots}^{R S \ldots},
\end{align}
with
\begin{align}\label{eq:spin-free-ops}
  &\text{with} \;\; \hat{E}^{P Q \ldots}_{RS \ldots} = \sum_{\substack{\sigma_{1} \sigma_{2} \ldots \\ \tau_{1} \tau_{2} \ldots}}^{\{\alpha,\beta\}} \hat{a}^{p_{\sigma_1 } q_{\sigma_{2}} \ldots}_{r_{\tau_{1}} s_{\tau_{2}} \ldots}
\end{align}
corresponding to the standard spin-free excitation operators.~\cite{helgakerMolecularElectronicstructureTheory2000}

Note that unlike spin-orbital coefficients, spin-free quantities are not antisymmetric under arbitrary index permutations; they are only symmetric with respect to simultaneous permutations of pairs of lower and corresponding upper indices comprising a column of indices
\begin{equation}\label{eq:spinfree-perm-sym}
  O_{RS \ldots}^{PQ \ldots} = O_{S R \ldots}^{QP \ldots}.
\end{equation}
The all-$\alpha$ component can be obtained by applying the antisymmetrizer of the symmetric group $S_{N}$ to the spin-free tensor
\begin{equation}\label{eq:full-alpha}
  o_{pq \ldots}^{rs \ldots} = \frac{1}{(N + 1)!} \sum_{\mathcal{P} \in S_{N}} (-1)^{\sigma} O^{\mathcal{P}(PQ \ldots)}_{RS \ldots}
\end{equation}
where $N$ is the number of upper (or lower) indices and $\sigma$ corresponds to the parity of the permutation $\mathcal{P}$.~\cite{shamasundarCumulantDecompositionReduced2009}
For example, for one-, two- and three-body operators, this yields the following relations
\begin{align}\label{eq:alphas-one-body}
  o_{r}^{p} &= \frac{1}{2} O_{R}^{P}, \\ \label{eq:alphas-two-body}
  o_{rs}^{pq} &= \frac{1}{6} \left( O_{RS}^{PQ} - O_{RS}^{QP} \right), \\
  o^{pqr}_{stu} &= \frac{1}{24} \big(
    O^{PQR}_{STU} - O^{PRQ}_{STU} - O^{QPR}_{STU} \\\notag
  &\qquad\quad - O^{RQP}_{STU} + O^{RPQ}_{STU} + O^{QRP}_{STU} \big).
\end{align}
By exploiting the property
\begin{equation}
  O^{PQR}_{STU} + O^{PRQ}_{STU} +  O^{QPR}_{STU} + O^{RQP}_{STU} + O^{RPQ}_{STU} + O^{QRP}_{STU} = 0,
\end{equation}
valid for RDMs and cumulants,~\cite{shamasundarCumulantDecompositionReduced2009} the expression for the three-body operator can be simplified further
\begin{equation}
  o^{pqr}_{stu} = \frac{1}{12} \left(O^{PQR}_{STU} + O^{RPQ}_{STU} + O^{QRP}_{STU}\right).
\end{equation}

Recall that, as demonstrated by Eq.~\eqref{eq:all-alphas}, these all-$\alpha$ quantities are redundant and can be expressed in term mixed $\alpha\beta$ spin components.
To express only the non-redundant components, \emph{partial-trace} relations~\cite{shamasundarCumulantDecompositionReduced2009} can be exploited, which state that integrating out a subset of the spin indices leaves the remaining spin-orbital coefficients satisfying the same spin relations as lower-body operators.
For instance, the partially traced operator containing a single pair of spin-orbital indices
\begin{align}\label{eq:partial-trace}
  O^{p_{\alpha} Q R \ldots}_{s_{\alpha} T U \ldots} \coloneqq \sum_{\substack{\sigma_1 \sigma_2 \ldots \\ \tau_1  \tau_2  \ldots}}^{\{\alpha,\beta\}} o^{p_{\alpha} q_{\sigma_{1}} r_{\sigma_{2}} \ldots}_{s_{\alpha} t_{\tau_1  } u_{\tau_2 } \ldots }
\end{align}
satisfies $O^{p_{\alpha} QR \ldots}_{s_{\alpha} TU} = O^{p_{\beta} QR \ldots}_{s_{\beta} TU}$, mirroring the one-body relation from Eq.~\eqref{eq:one-body-beta2alpha}.
Using the partial trace relations, the non-redundant spin component can be expressed entirely in terms of spin-free quantities, as illustrated here for a two-body operator
\begin{align}
  o^{pq}_{rs} + o^{p \overline{q}}_{r \overline{s}} &\overset{\eqref{eq:partial-trace}}{\eqqcolon} O^{p Q}_{r S}
  \overset{\eqref{eq:alphas-one-body}}{=} \frac{1}{2} O^{P Q}_{R S}, \\
  \leftrightarrow \quad o^{p \overline{q}}_{r \overline{s}} &= \frac{1}{2} O^{P Q}_{R S} - o^{pq}_{rs}, \\
  \overset{\eqref{eq:alphas-two-body}}{\leftrightarrow} \quad o^{p \overline{q}}_{r \overline{s}} &= \frac{1}{2} O^{P Q}_{R S} - \frac{1}{6} ( O^{P Q}_{R S} - O^{Q P}_{S R} ), \\
  \leftrightarrow \quad o^{p \overline{q}}_{r \overline{s}} &= \frac{1}{6} \left( 2 O^{P Q}_{R S} + O^{Q P}_{S R} \right).
\end{align}
The same approach can be generalized to arbitrary $n$-body operators, using partial traces to connect non-redundant components to lower-body relations.
Explicit formulas for up to four-body operators can be found in Ref.~\citenum{shamasundarCumulantDecompositionReduced2009}.

It should be stressed that all of the above relations are only formally valid for singlet operators.
Therefore, to make use of these relations within our approach, it is necessary to demonstrate that all tensors involved in the contractions from the generalized Wick's theorem are indeed singlets.
These tensors consist of one- and two-electron integrals, cluster amplitudes, one-particle and one-hole RDMs, and higher-order cumulants.
The integrals stem from a spin-adapted CAS self-consistent field (CASSCF) calculation that relies on a restricted set of molecular orbitals, guaranteeing therefore the singlet nature of the integrals.
Since the MRCC wave function should not alter the spin of the reference CASSCF solution, the cluster amplitudes must act as singlets in order to conserve this property.
The remaining question is why the RDMs and cumulants derived from a CASSCF wave function behave as singlets, even when the reference state can exhibit arbitrary spin multiplicity---an issue examined in the following section.

\subsubsection{Extensions to Spin Multiplets}

In principle, the derivation presented so far is not immediately applicable to reference states with higher than singlet spin multiplicities.
One reason for this restriction is that spin-orbital RDMs generally depend on the spin projection quantum number $M_{S}$ and, consequently, lack invariance under spin rotations.
As they are defined in terms of expectation values, however, their spin-dependence can be simply integrated out as in Eq.~\eqref{eq:spinfree-spinsum}.
The resulting spin-free quantity is identical for all states within the multiplet~\cite{kutzelniggCumulantExpansionReduced1999}, a property that can be demonstrated through the Wigner-Eckart therorem.~\cite{shamasundarCumulantDecompositionReduced2009}
Another reason is that, unlike RDMs, cumulants are not defined in terms of expectation values (see Eq.~\eqref{eq:cumulant}) and, therefore, performing simple spin summation yields quantities that remain $M_{S}$-dependent.~\cite{kutzelniggCumulantExpansionReduced1999}
Several strategies have been proposed to eliminate this dependency.~\cite{kutzelniggCumulantExpansionReduced1999}
The approach adopted in this work relies on the observation that, for any spin multiplicity, an $M_{S}$-invariant state can be constructed by forming an equally weighted ensemble average over all members of the multiplet
\begin{equation}
  \rho(S) = \frac{1}{2 S + 1} \sum_{M_{S} = -S}^{S} \ket{\Psi(S, M_{S})}\bra{\Psi(S, M_{S})},
\end{equation}
resulting in a state that is a singlet.
Here, $S$ denotes the total spin quantum number.
The spin- and spatial orbital RDMs can be extracted from $\rho(S)$ by taking the trace
\begin{align}\label{eq:spin-orbital-rdm}
  \gamma^{pq \ldots}_{rs \ldots}(S) &= \operatorname{tr}\left(\rho(S) \, \hat{a}^{pq \ldots}_{rs \ldots}\right), \\\label{eq:spinfree-rdm}
  \Gamma^{PQ \ldots}_{RS \ldots} &= \operatorname{tr}\left(\rho(S) \, \hat{E}^{PQ \ldots}_{RS \ldots}\right).
\end{align}
Performing spin integration of the density cumulants, using the spin-orbital RDMs defined in Eq.~\eqref{eq:spin-orbital-rdm} results in expressions that are $M_{S}$-independent as they can be written entirely in terms of spin-free RDMs from Eq.~\eqref{eq:spinfree-rdm}.\cite{kutzelniggCumulantExpansionReduced1999,shamasundarCumulantDecompositionReduced2009,kutzelniggSpinfreeFormulationReduced2010}

One might initially assume that constructing the ensemble-averaged density matrix $\rho(S)$ and corresponding RDMs requires solving for every state in the multiplet---a task that becomes increasingly expensive for higher multiplicities.
For instance, in the context of DSRG, Li and Evengelista evaluate the ensemble-averaged RDMs by, first, computing the high-spin $M_{S} = S$ state and, then, successively applying the lowering operator $\hat{S}_{-}$ to recover the lower spin states in the multiplet.~\cite{liSpinfreeFormulationMultireference2021}
Since the final spin-adapted equations only require spin-free RDMs, this procedure can be avoided by exploiting the linearity of the trace and recalling that any spin-independent quantity is identical, by definition, for each state in the multiplet and, hence, also for the ensemble-averaged state as a whole
{\footnotesize
  \begin{align}\notag
    \operatorname{tr}&\left(\rho(S) \, \hat{E}^{PQ \ldots}_{RS \ldots}\right) \\
    &= \frac{1}{2 S + 1} \sum_{M_{S} = -S}^{S} \operatorname{tr}\!\left(\ket{\Psi(S, M_{S})}\bra{\Psi(S, M_{S})} \, \hat{E}^{PQ \ldots}_{RS \ldots}\right) \\
    &= \frac{1}{2 S + 1} \sum_{M_{S} = -S}^{S} \Gamma_{RS \ldots}^{PQ \ldots} = \Gamma_{RS \ldots}^{PQ \ldots}.
  \end{align}
}
Therefore, if the spin-free RDMs are available from a spin-adapted CAS solver, as in the case of the ORCA program,~\cite{Neese2020Jun} they can directly be used to compute the corresponding spin-free cumulants.
Otherwise, if only a CAS solver in spin-orbital basis is available, it is sufficient to resolve one of the members of the multiplet and integrate out the spin degrees of freedom to obtain the corresponding spin-free RDMs.~\cite{Kutzelnigg1982Sep}

\section{Implementation}\label{sec:implementation}

Our implementation of the spin-free RIC-MRCCSD equations employs a combination of automatic code generation tools.
ORCA's native code generator \texttt{AGE}~\cite{lechnerCodeGenerationORCA2024} is capable of deriving the projected form of the residual equations.
Therefore, Evangelista's \texttt{Wick\&d} program~\cite{evangelistaAutomaticDerivationManybody2022} is needed instead to produce the many-body residual equations in spin-orbital basis.
These equations are then processed by \texttt{AGE} , which performs the spin adaptation and produces the C++ code for evaluating the contractions.
In the process, \texttt{AGE} further optimizes the resulting equations by factorizing them into binary tensor contractions and identifying reusable intermediates across contractions.~\cite{lechnerCodeGenerationORCA2024}

\subsection{\texttt{Wick\&d} to \texttt{AGE} Translator}

Once \texttt{Wick\&d} generates the equations in its internal representation, it can output the resulting tensor contractions in several formats, including the familiar \texttt{NumPy}~\cite{harris2020array} optimized \texttt{Einsum}~\cite{daniel2018opt} expressions, which the pilot implementation relied on~\cite{feldmannRenormalizedInternallyContracted2024}.
In the present study, a local version of \texttt{Wick\&d} has been extended to emit the tensor contractions directly in an \texttt{AGE} compatible syntax.

An important subtlety in using \texttt{Wick\&d} is that it does not automatically enforce the antisymmetry of the residuals inherited from their corresponding amplitudes.
Instead, it computes the non-symmetric tensor $g_{p_{\sigma_1 }q_{\sigma_2 }}^{r_{\sigma_3}s_{\sigma_4}}$ from which the residual may be obtained by antisymmetrizing the indices pertaining to the same orbital space.
For example, in the case of a two-hole two-particle (2h2p) excitation, the contributions to the residual $r_{i_{\tau_1  }j_{\tau_2  }}^{a_{\sigma_1  }b_{\sigma_2  }}$ should be antisymmetrized as
\begin{equation}\label{eq:antisym}
  r_{i_{\tau_1  }j_{\tau_2  }}^{a_{\sigma_1   }b_{\sigma_2   }} \leftarrow g_{i_{\tau_1  }j_{\tau_2  }}^{a_{\sigma_1 }b_{\sigma_2   }} - g_{j_{\tau_2 }i_{\tau_1  }}^{a_{\sigma_1   }b_{\sigma_2   }} - g_{i_{\tau_1 }j_{\tau_2  }}^{b_{\sigma_2   }a_{\sigma_1   }} + g_{j_{\tau_2 }i_{\tau_1  }}^{b_{\sigma_2   }a_{\sigma_1  }}.
\end{equation}
This design is driven by efficiency consideration, since antisymmetrizing the single tensor $g_{p_{\sigma_1 }q_{\sigma_2 }}^{r_{\sigma_3}s_{\sigma_4}}$  is usually computationally less expensive than repeatedly evaluating the contractions with permuted indices in order to recover the other three contributions.
Our spin-adapted implementation, however, only requires the mixed $\alpha\beta$ spin sectors for two particle excitations.
All of these contributions are then directly accumulated into the residual $r_{i \overline{j}}^{a \overline{b}}$.

\subsection{Excitation Classes}

\begin{figure*}[htb]
  \centering
  \includegraphics[width=0.7\textwidth]{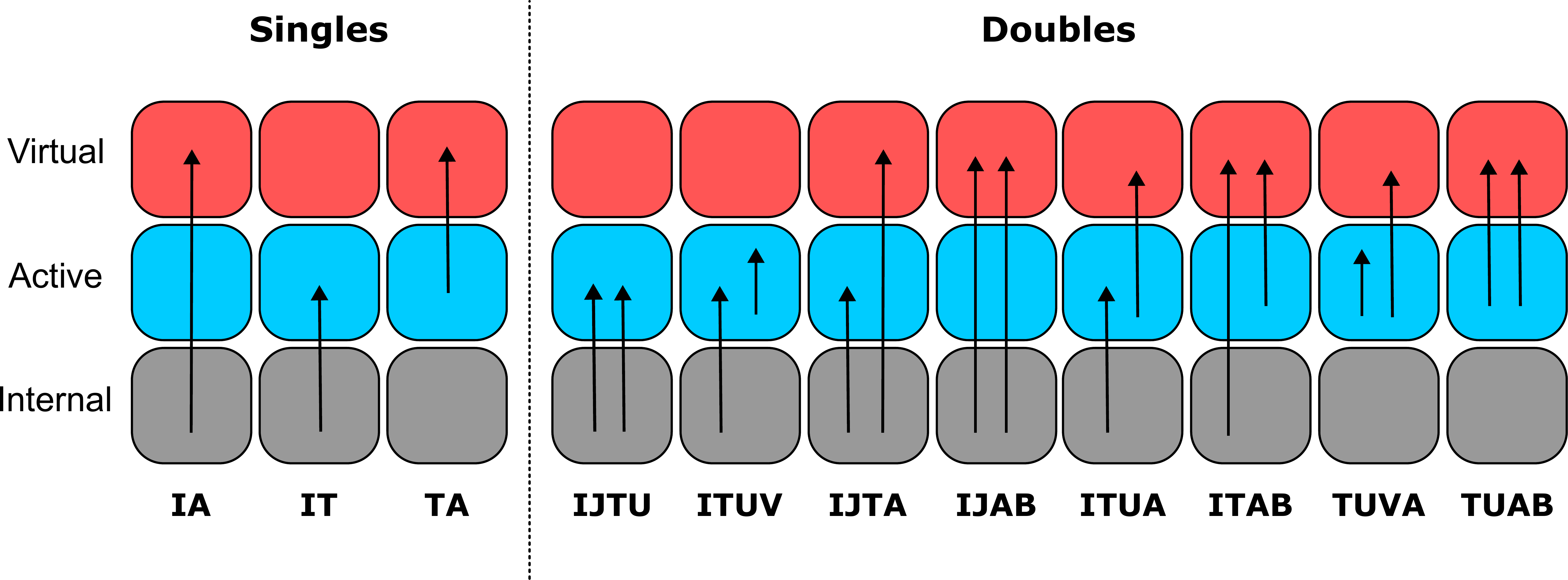}
  \caption{
    Breakdown of the amplitude excitation classes present in the RIC-MRCCSD scheme.
    Each column corresponds to a specific excitation class with each arrow representing a single-electron excitation.
    The name of the excitation class is found under each excitation with the first (two) and last (two) letters denoting the hole and particle spaces for the singles (doubles) excitations, respectively.
  }
  \label{fig:exc-classes}
\end{figure*}

In ORCA, all tensors involved in the contractions---including amplitudes---are decomposed by \emph{excitation class}, defined by the orbital spaces of the indices.
For single-particle excitations, this results in three distinct classes with corresponding amplitudes $t_{I}^{A}$ (as in single-reference schemes), $t_{I}^{T}$, and $t_{T}^{A}$.
For the doubles, the excitations fall into eight classes, analogous to the \emph{first-order interacting spaces} in conventional IC-MR
schemes.~\cite{McLean1973Feb}
Figure~\ref{fig:exc-classes} depicts all the distinct amplitude excitation classes.

Each excitation class has a designated tensor that stores the coefficients for that class.
To apply the singlet-constraining relations for doubles excitation that relate, for instance, $t^{p_\alpha q_\alpha}_{r_{\alpha} s_{\alpha}}$ to $t^{p_{\alpha} q_{\beta}}_{r_{\alpha} s_{\beta}}$ components, one must permute either the upper or lower indices of the tensor.
In principle, either choice is valid.
However, for some of the excitation classes, the two particle and hole spaces might differ from one another and, therefore, cannot be permuted.
For instance, the $t_{IJ}^{TA}$ class excites two electrons from the internal space into an active and virtual orbital.
In this case, the particle indices lie in different spaces, so they cannot be permuted; only the hole indices can be permuted---both being internal.
Taking this into account, the following scheme is used in our implementation:

\begin{itemize}
  \item For $t_{IT}^{UV}$ and $t_{IT}^{AB}$, the particle indices are permuted.
  \item For $t_{IJ}^{JA}$ and $t_{TU}^{VA}$, the hole indices are permuted.
  \item For $t_{IJ}^{AB}$, $t_{IJ}^{TU}$, and $t_{TU}^{AB}$, both particle and hole indices are in the same space, allowing either one to be permuted.
    We choose to permute the upper indices.
    Additionally, for these three classes, we exploit the permutational symmetry in Eq.~\eqref{eq:spinfree-perm-sym} to halve the storage requirements of these tensors.
  \item The $t_{IT}^{AU}$ class is special: both hole and particle indices lie in different spaces, so neither set can be permuted.
    Therefore, for this class, we store both $t_{I\overline{T}}^{U\overline{A}}$ and $t_{I\overline{T}}^{\overline{U}A}$ in order apply the singlet constraining relation
    \begin{equation}
      t_{IT}^{UA} = t_{I\bar{T}}^{U\overline{A}} + t_{I\overline{T}}^{\overline{U}A}
    \end{equation}
\end{itemize}

Note that these considerations are only needed for the amplitude tensors and not for the cumulants, since their indices lie entirely in the active space and can therefore be freely permuted.
These singlet constraining relations alongside their redundant and non-redundant spin components for the excitation classes are summarized in Table~\ref{tab:doubles-singlet-relation}.

\begin{table*}[htb]
  \centering
  \begin{tabular}{@{} l c c @{}}
    \toprule
    Class & Non-Redundant & Relation \\\midrule
    IA, IT, TA, TU & $o_{q}^{p}$ & $o_{\overline{q}}^{\overline{p}} = o_{q}^{p}$ \\\\
    ITUV, ITAB, (IJAB, IJTU, TUAB, TUVW) & $o^{p \overline{q}}_{r \overline{s}}$ & $o^{pq}_{rs} = o^{p \overline{q}}_{r \overline{s}} - o^{p \overline{q}}_{s \overline{r}}$ \\
    IJTA, TUVA, (IJAB, IJTU, TUAB, TUVW) & $o^{p \overline{q}}_{r \overline{s}}$ & $o^{pq}_{rs} = o^{p \overline{q}}_{r \overline{s}} - o^{q \overline{p}}_{r \overline{s}}$ \\
    ITUA & $o^{p \overline{q}}_{r \overline{s}}$, $o^{\overline{p}q}_{r\overline{s}}$ & $o^{pq}_{rs} = o^{p\overline{q}}_{r\overline{s}} + o^{\overline{p}q}_{r\overline{s}}$ \\\\
    TUVXYZ & $o^{tu \overline{v}}_{xy \overline{z}}$ & $o^{tuv}_{xyz} = o^{tu \overline{v}}_{xy \overline{z}} - o^{tv \overline{u}}_{xyz} - o^{v u \overline{t}}_{xyz}$ \\
    \bottomrule
  \end{tabular}
  \caption{
    Singlet-constraining relations summarizing non-redundant spin components for the amplitudes and density cumulants and their relations for deriving redundant components.
    Due to the decomposition of the tensors in terms of orbital spaces of the indices, only some of the relations can be used.
    Classes for which multiple rules can be applied are grouped in parentheses.
    The tensor $o$ can symbolize amplitudes or density cumulants.
  }
  \label{tab:doubles-singlet-relation}
\end{table*}

\subsection{Spin-Adaptation Procedure}
\begin{figure*}[htb]
  \centering
  \includegraphics[width=1.0\textwidth]{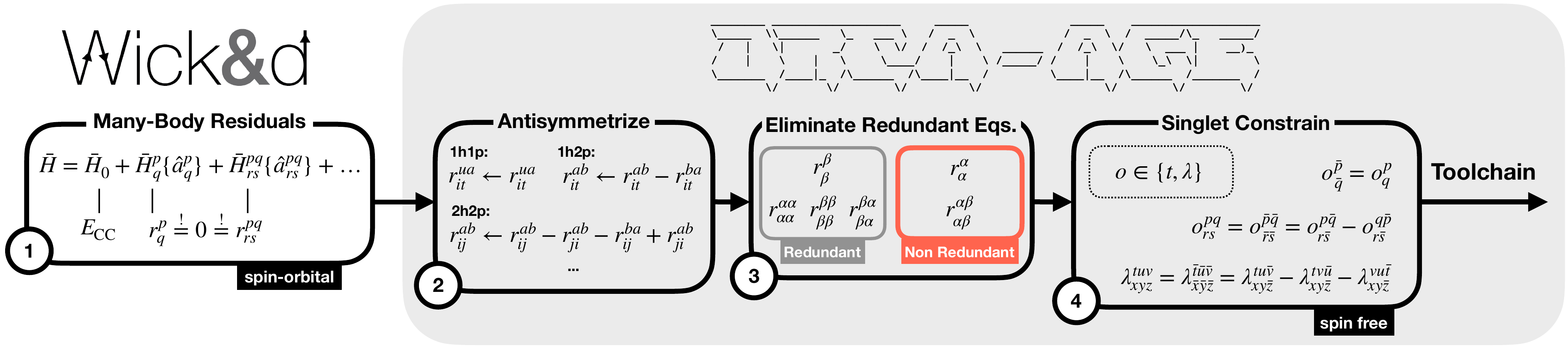}
  \caption{
    Procedure for deriving spin-free many-body residual equations.
    The pipeline begins with spin-orbital residuals from \texttt{Wick\&d}, applies antisymmetrization, discards redundant spin sectors exploits singlet constraints and produces non-redundant spin-free equations, which are then passed to the remaining \texttt{AGE} toolchain for C++ code generation.
    The notation $n$h$m$p in the antisymmetrization step (2) denotes the excitation class producing $n$ holes and $m$ particles.
  }
  \label{fig:spin-adapt-pipeline}
\end{figure*}

With this theoretical framework in place, this section presents the complete procedure for spin adapting the many-body residual equations, with the overall workflow depicted in Figure~\ref{fig:spin-adapt-pipeline}.

Starting from the energy and residual equations produced by \texttt{Wick\&d} in spin-orbital form, first, antisymmetrization is applied to the produced residuals as in Eq.~\eqref{eq:antisym}.
Then, explicit spin indices are introduced to the tensors, allowing residual equations pertaining to redundant spin sectors to be discarded, as these are not used to update non-redundant amplitudes.
The tensors---namely the amplitudes and cumulants---involved in the remaining contractions are subsequently expressed in terms of non-redundant quantities via the singlet-constraining relations.
Recall that these blocks are $o_{\alpha}^{\alpha}$, $o_{\alpha\beta}^{\alpha\beta}$, and $o_{\alpha\alpha\beta}^{\alpha\alpha\beta}$ for the one-, two-, and three-body operators, respectively.
Each of these quantities can be cast in terms of their spin-free analogs, yielding equations in which spin dependence is entirely eliminated.
These spin-free equations are then passed on to the remaining \texttt{AGE} toolchain, which optimizes the tensor contractions and generates high-performance, MPI-parallelized C++ code to execute them.

\section{Results}\label{sec:results}

In this section, we present a detailed benchmarking study of the RIC-MRCCSD scheme across a wide range of molecular systems.
The primary goal is to validate key properties such as size consistency and to assess computational efficiency and accuracy in comparison to established electronic structure methods.

\subsection{Computational Methodology}
To evaluate the impact of spin adaptation, the execution times against both restricted and unrestricted single-reference coupled cluster methods are evaluated.
These serve as the natural single-reference analogs.
The accuracy and performance of RIC-MRCCSD are further assessed relative to prominent multireference approaches, including MR perturbation theory (MRPT), MR configuration interaction (MRCI), and MRCC---each expressed using the fully internally-contracted formalism.
Our MRPT calculations considered the widely used $n$-electron valence state perturbation theory of second order (NEVPT2)~\cite{Angeli2001Jun} as well as a recent approximate fourth-order variant~\cite{kempferEfficientImplementationApproximate2025} developed by some of the authors of the present work.
In addition, the zeroth-order MR coupled-electron approximation method (CEPA(0)) with singles and doubles excitations is used as an approximation to the MRCISD equations, because
it has demonstrated high accuracy in benchmark studies, frequently surpassing that of the latter~\cite{blackLinearQuadraticInternally2019,Waigum2025Jan}.
As the most accurate but computationally demanding method, the MRCCSD~\cite{lechnerCodeGenerationORCA2024}) scheme implemented in ORCA was considered.
This approach corresponds to the icMRCCSD-A scheme from the work of Hanauer and K\"ohn.~\cite{hanauerPilotApplicationsInternally2011}

Except for the NEVPT2 implementation, which was manually optimized~\cite{kollmarEfficientImplementationNEVPT22021}, all working equations for these methods were derived using the AGE automatic code generator from ORCA~\cite{lechnerCodeGenerationORCA2024}.
All calculations were performed with a development version of ORCA 6.1.~\cite{Neese2020Jun}
Unless stated otherwise, timing benchmarks were carried out on compute nodes equipped with two 12-core Intel(R) Xeon(R) E5-2687W v4 CPUs.

\subsection{Size Consistency}

In this section, the size consistency of the RIC-MRCCSD method is verified numerically.
A method will be considered size-consistent if the total energy of two non-interacting subsystems equals the sum of the energies of the individual subsystems.
This property is critical for producing reliable energies across different molecular structures and is a key reason for the success of coupled cluster methods over configuration interaction approaches.

To probe size consistency, calculations were carried out on systems composed of three monomers---ethene, butadiene, and hexatriene---using an active space corresponding to each monomer's $\pi$-system.
Pairs of monomers were placed 100 {\AA}ngstr\"om apart to eliminate all interactions between them, and the energy of the combined system was compared to the sum of the energies computed for each monomer individually.
For a size-consistent method, these energies should be identical.
The size consistency error for two monomers $i$ and $j$, $\Delta E_{\text{sc}}$, is, hence, defined as
\begin{equation}
\Delta E_{\text{sc}} = E(\text{mon}_i + \text{mon}_j) - E(\text{mon}_i) - E(\text{mon}_j).
\end{equation}

Calculations were carried out with the def2-SVP~\cite{weigendBalancedBasisSets2005} atomic orbital basis set and a tight energy convergence criterion of $1\times10^{-14}$ [$E_{h}$] for the CASSCF solution.

\begin{table*}[htb]
\centering
\caption{Size consistency errors reported in $E_h$ for pairs of the monomers: ethene, butadiene and hexatriene.}
\label{tab:size-consistency}
\begin{tabular}{@{}l r r@{}}
  \toprule
  \textbf{Molecules} & $\Delta E_{\text{sc}}(\text{CASSCF})$ & $\Delta E_{\text{sc}}(\text{RIC-MRCCSD})$ \\
  \midrule
  2 $\times$ Ethene           &  $4.7 \times 10^{-8}$          &   $8.4 \times 10^{-8}$    \\
  2 $\times$ Butadiene        &  $- 7.3 \times 10^{-10}$                    & $-6.2 \times 10^{-10}$    \\
  2 $\times$ Hexatriene       &  $4.8 \times 10^{-10}$          &  $3.0 \times 10^{-10}$ \\
  Ethene + Butadiene   &  $-7.2 \times 10^{-11}$                    & $-1.3 \times 10^{-9}$    \\
  Ethene + Hexatriene  &  $-8.9 \times 10^{-11}$                    & $-2.6 \times 10^{-9}$    \\
  \bottomrule
\end{tabular}
\end{table*}

Table~\ref{tab:size-consistency}
reports the size-consistency errors for both the CASSCF and RIC-MRCCSD solutions.
All errors are within $10^{-8}$ [Eh] or smaller, with RIC-MRCCSD deviations comparable to those from CASSCF.
Given that CASSCF is rigorously size-consistent, the small deviations observed can be attributed to numerical noise.

\subsection{Comparison with Single-Reference Methods}

\begin{figure}[htb]
\centering
\includegraphics[width=0.5\linewidth]{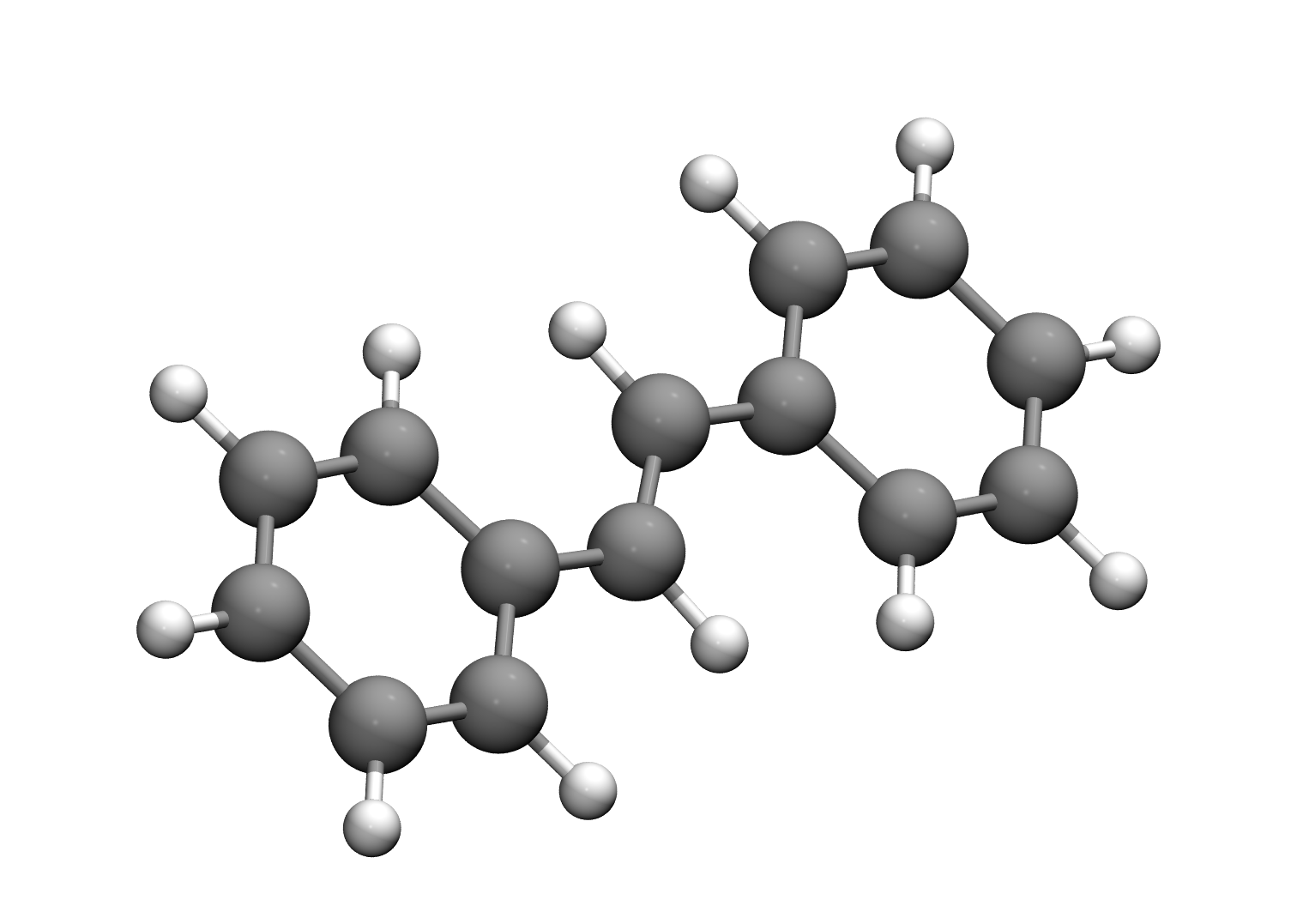}
\caption{Trans-stilbene molecular structure (hydrogen atoms in white, carbon atoms in gray).}
\label{fig:stilbene-struct}
\end{figure}

This section assesses the efficiency of the spin-free implementation of the RIC-MRCCSD by benchmarking its performance against conventional single-reference coupled cluster schemes.
For this comparison, the trans-stilbene molecule shown in Figure~\ref{fig:stilbene-struct} was selected.
This system features a fairly large active space of CAS(14, 14) comprised of the conjugated $\pi$-system spanning the two benzene rings and the two bridging carbon atoms.

Single-iteration runtimes of the RIC-MRCCSD method are compared to those from both restricted and unrestricted SR-CCSD approaches.
Before discussing the results, we briefly summarize the key differences in terms of the equations present in the two formalisms.
SR schemes lack an active space containing partially occupied orbitals These orbitals have the particularity that electrons can be excited to and from---unlike the internal and virtual space.
This leads to a huge number of additional equations in MR schemes, of which only a subset is preserved in RIC-MRCCSD for computational reasons (see above).
In principle, the two methods share the contractions involving only internal and virtual indices.
However, since RIC-MRCCSD truncates the BCH expansion at the the second nested commutator, contractions arising from the third and fourth commutator---present in SR-CCSD--are omitted.
Therefore, RIC-MRCCSD does not reduce to a regularized version of SR-CCSD in the limit of a vanishing active space (CAS(0,0)).

To summarize, RIC-MRCCSD both introduces additional contractions due to the active space and omits certain contractions found in standard SR-CCSD.
Nevertheless, after factorization, both approaches share the same rate-limiting contraction, as described in Ref.~\citenum{feldmannRenormalizedInternallyContracted2024}, and therefore exhibit the same formal computational scaling.

\begin{table}[htb]
\centering
\begin{tabular}{l cc c}
  \toprule
  & \multicolumn{2}{c}{CCSD} &  \\ \cmidrule(lr){2-3}
  Processes  & RHF    & UHF    &  RIC-MRCCSD       \\
  \midrule
  1  & 3590 & 18228 & 8580 \\
  2  & 2054 & 10240 & 5353 \\
  4  & 1224 & 6319   & 3711 \\
  8  & 838  & 4430  & 2145 \\
  16 & 743  & 3267  & 1869 \\
  \bottomrule
\end{tabular}
\caption{
  Parallel runtimes, in seconds, of a single iteration of various coupled cluster methods for the trans-stilbene molecule in the def2-TZVP bases.
  The RIC-MRCCSD scheme employs an active space of CAS(14, 14).
}
\label{tab:stilbene-timings}
\end{table}
Table~\ref{tab:stilbene-timings} summarizes the execution times for a single iteration of each coupled cluster method, covering both serial and parallel MPI runs with 2 to 16 processes on a single node.
These data are based on
calculations with the def2-TZVP~\cite{weigendBalancedBasisSets2005} basis set and the frozen-core approximation, keeping core orbitals doubly occupied throughout and thereby excluding excitations from these orbitals.

\begin{figure}[htb]
\centering
\includegraphics[width=\linewidth]{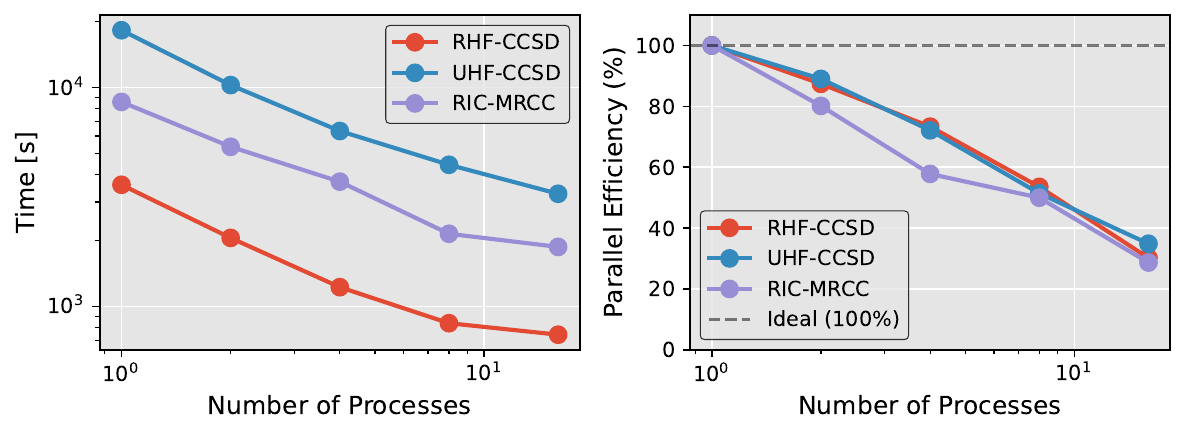}
\caption{
  Execution time and parallel efficiency of a single iteration of various coupled cluster methods for trans-stilbene with a def2-TZVP basis and an active space of CAS(14, 14) for the RIC-MRCCSD method.
}
\label{fig:stilbene-timings}
\end{figure}

Our RIC-MRCCSD implementation demonstrates competitive performance compared to SR-CCSD methods.
Each RIC-MRCCSD iteration is less than 3.5 times slower than RHF-CCSD, despite containing a significantly larger number of contractions.
The data also highlight the efficiency gains from the spin adaptation as RIC-MRCCSD is substantially faster than the conventional unrestricted CCSD formalism.
Figure~\ref{fig:stilbene-timings} presents the timing data and the corresponding parallel efficiency,
\begin{equation}
\text{Parallel Efficiency} = \frac{\text{Speedup}}{\text{\# Processes}},
\end{equation}
as the number of MPI processes increases.
Although parallel efficiency declines with the number of processes, significant speed-ups are still achieved, particularly with fewer processes.
RIC-MRCCSD appears to exhibit slightly lower parallel efficiency with two and four processes, but overall, its scaling closely matches that of the single-reference schemes.

\subsection{Scaling with Molecular Size}

This section examines the computational scaling of the RIC-MRCCSD method using the all-\emph{E} series of polyenes (2 to 14 carbon atoms) and corresponding active spaces with the def2-SVP~\cite{weigendBalancedBasisSets2005} basis set.

\begin{figure*}[htb]
\centering
\includegraphics[width=\linewidth]{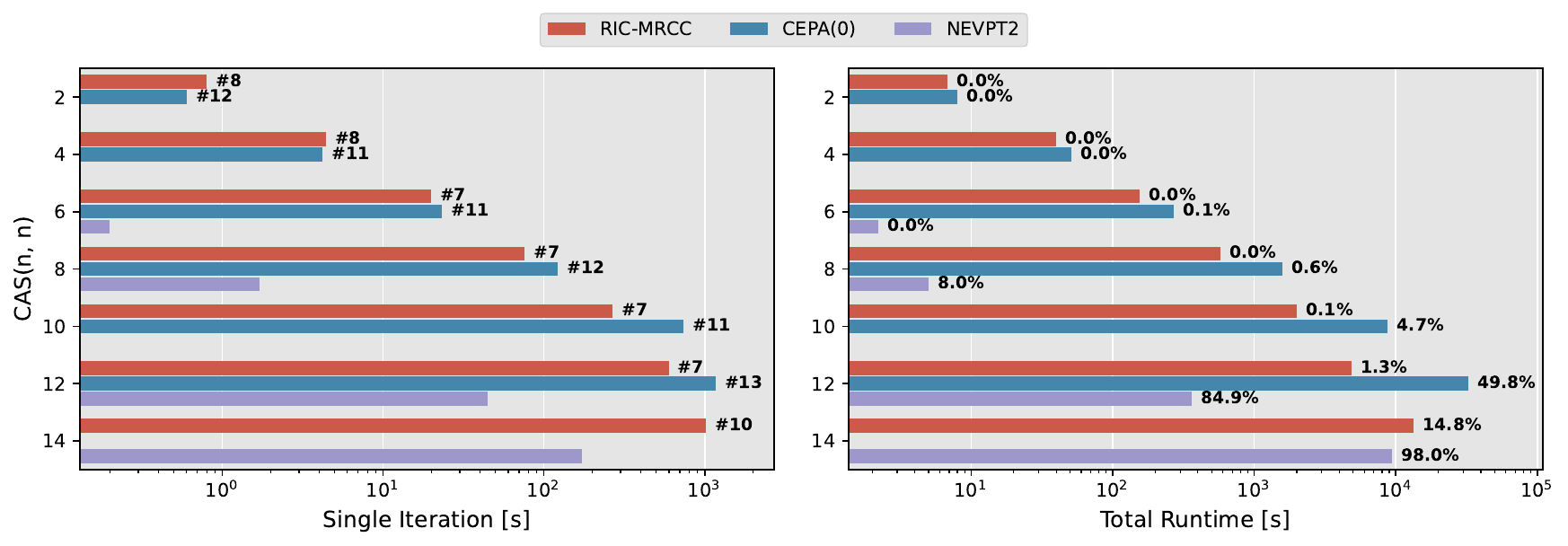}
\caption{
  Runtimes for various multireference methods for an all-E polyene series (2 to 14 carbon atoms) with corresponding active spaces defined by the conjugated $\pi$-system.
  Left: timings per iteration with the number of iterations to convergence indicated next to each bar (except for NEVPT2 which is non-iterative).
  Right: total runtimes for each method and the percentage of the total time taken for computing the 1-, 2- and 3-RDMs for RIC-MRCCSD (and the additional 4-RDM required for NEVPT2 and CEPA(0)).
  Note that for the largest polyene, CEPA(0) failed to converge and is therefore absent from the plots.
}
\label{fig:polyenes-timings}
\end{figure*}

Figure~\ref{fig:polyenes-timings} (left panel) shows the runtime per iteration for RIC-MRCCSD, CEPA(0), and NEVPT2 (which is assumed to converge in a single iteration), while the right panel reports the total execution times.
Additionally, the number of iterations required for convergence and the fraction of the total runtime devoted to the RDM computations are indicated next to each bar.
Initially, RIC-MRCCSD shows slower per-iteration runtimes than CEPA(0) for the smallest systems.
However, this does not lead to longer total execution times, as RIC-MRCCSD consistently converges in 30\% to 50\% fewer iterations.
In addition, RIC-MRCCSD shows significantly better scaling with active space size: for all polyenes larger than butadiene (CAS(4,4)), each iteration is faster than CEPA(0), and the performance gap widens as system size increases.
Moreover, the total runtime also scales more favorable with RIC-MRCCSD, as it does not require the evaluation of the 4-RDM.
This tends to be a significant computational bottleneck in most schemes as illustrated by the CAS(12, 12) CEPA(0) calculation where almost half of the total runtime is spent on computing this quantity.

When comparing to NEVPT2 for the smaller systems, RIC-MRCCSD is not competitive in terms of time to solution, where NEVPT2 is substantially faster.
It is encouraging, however, to see that once an active space size of CAS(14, 14) is reached, RIC-MRCCSD is only 40\% slower than the highly optimized implementation of NEVPT2 in ORCA~\cite{guoApproximationsDensityMatrices2021a,kollmarEfficientImplementationNEVPT22021}.
Unlike CEPA(0), which evaluates the 4-RDM explicitly, the NEVPT2 implementation avoids this by directly computing the contribution of the 4-RDM to its equations~\cite{kollmarEfficientImplementationNEVPT22021}.
Despite this, this contribution still accounts for 98\% of NEVPT2's total runtime, which explains the narrowing performance gap between RIC-MRCCSD and NEVPT2 for larger active spaces.

\subsection{Transition-Metal Ion Excitation Energies}

This section assesses the accuracy of the RIC-MRCCSD method by comparing its state-averaged excitation energy errors to those from other multireference approaches, using a benchmark set of transition-metal ions~\cite{Neese2007Feb,Reimann2023Jan}.
The benchmark consists of 56 excitation energies for seven divalent and seven trivalent fourth-row transition-metal ions, all calculated with a DKH-def2-QZVPP basis set~\cite{weigendBalancedBasisSets2005}.
The energies are evaluated against experimental values found in the NIST database.~\cite{Kramida2024}
These states are averaged over the $J$ quantum number as outlined in the supporting information of Ref.~\cite{kempferEfficientImplementationApproximate2025}.
The supporting information also contains detailed electronic state assignments.
The method was assessed on each system using a smaller active space containing only the $3d$ orbitals and a larger active space including also the $4d$ orbitals in order to account for the double-shell effect.
For the copper ion, the $4s$ orbitals were also taken into the active space.

Note that our initial value of $s=0.5\;E_{h}^{-2}$ failed to converge with the larger active space for most states of the Co(III), Cu(III) and Ni(III) metals.
Such behavior was also observed by Li and Evangelista in their spin-free implementation of the sequentially-transformed DSRG when studying iron-water and iron-ammonium clusters.~\cite{liSpinfreeFormulationMultireference2021}
Their solution was to decrease the flow parameter to $s=0.1\; E_{h}^{-2}$, which increases the regularization and should improve numerical robustness.
This solution also resolved our convergence issues, with all calculations converging successfully for $s \le 0.4\; E_{h}^{-2}$.
It is important to note, however, thatRIC-MRCCSD iterations do not diverge for these systems.
Instead, they converge exceedingly slowly as $s$ is increased.
For instance, the ground-state of Ni(III) required 11, 18, 31 and 71 iterations to converge for $s=0.1, 0.2, 0.3$ and $0.4\; E_{h}^{-2}$, respectively.

\begin{figure*}
\centering
\includegraphics[width=\linewidth]{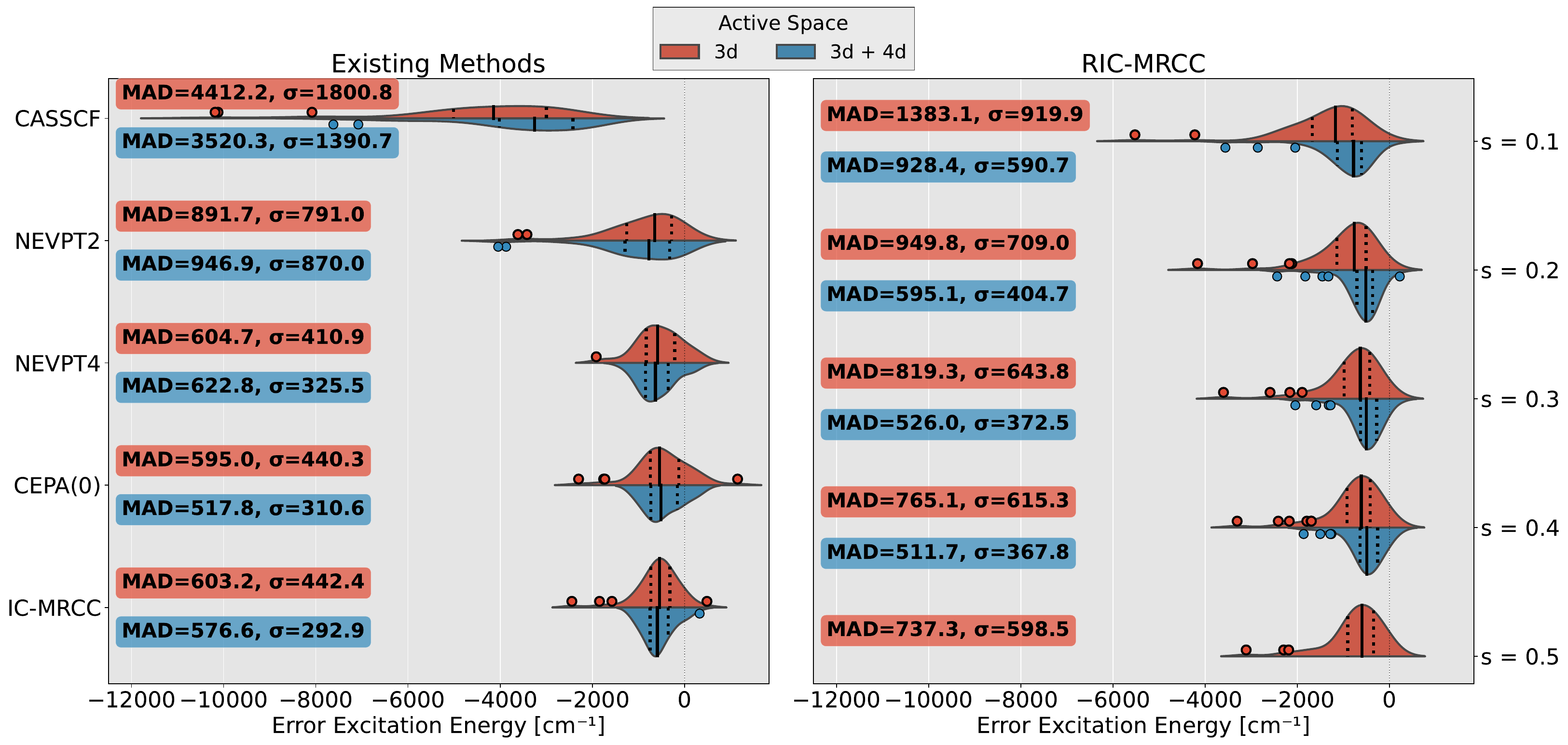}
\caption{
  Errors in excitation energies relative to experimental data are shown for a benchmark set of 14 2+ and 3+ transition-metal ions.
  Results are evaluated using two active space sizes comprised of the $3d$ orbitals (red) and $3d + 4d$ orbitals (blue), shown as separate halves of a violin plot.
  Additionally, the solid line represents the mean, the dotted lines indicate the first and third quartiles, and outliers are also highlighted.
  The left plot reports data from Ref.~\citenum{kempferEfficientImplementationApproximate2025} with IC-MRCC referring to the projection-based IC-MRCCSD formalism from K\"ohn, while the right plot presents new results for RIC-MRCCSD using various flow parameter values.
  Note that some RIC-MRCCSD calculations for the larger active space with $s=0.5 \; E_{h}^{-2}$ did not converge and are omitted.
  Each plot is complemented by its mean absolute difference (MAD) and standard deviation ($\sigma$).
}
\label{fig:tm-3d-4d}
\end{figure*}

Figure~\ref{fig:tm-3d-4d} displays error distributions in excitation energies for both active space sizes.
The left panel reports results from Ref.~\citenum{kempferEfficientImplementationApproximate2025} for standard multireference methods while the right panel presents RIC-MRCCSD results for different flow parameter values.
A breakdown of the errors for each electronic state can be found in the supporting information.
The results in Figure~\ref{fig:tm-3d-4d} show that increasing the flow parameter $s$ consistently reduces the error, as reflected in both the MAD and $\sigma$ statistics.
This trend is expected, since smaller values of $s$ recover more dynamic correlation and, hence, generally improve accuracy, but at the expense of numerical robustness (seen in the non-convergent cases for $s = 0.5\; E_{h}^{-2}$).
For the largest value still permitting convergence ($s = 0.4 \; E_{h}^{-2}$), the method's accuracy with the $3d$-only active space is somewhere between that of NEVPT2 and NEVPT4.
Including the $4d$ orbitals in the active space further reduces the error, with RIC-MRCCSD even outperforming IC-MRCC in terms of MAD.
However, the conclusion from this observation is somewhat nuanced.
The double-shell effect is traditionally invoked to compensate for missing dynamic correlation and should ideally not be necessary for high-accuracy MR methods.
Indeed, this trend is observed for most of the standard MR methods, which barely change as the active space is expanded.
As noted in Ref.~\citenum{kempferEfficientImplementationApproximate2025}, this is less conclusive for the NEVPT family of methods, where the energy actually increases with the larger active space, suggesting that these methods benefit from some fortuitous error cancellation.

\subsection{Ethylene Rotation}

This section evaluates the accuracy of the RIC-MRCCSD method in comparison to established multireference approaches by examining the dihedral rotation of ethylene.
The system is modeled using a CAS(2, 2) active space, consisting of the $\pi$-bonding and $\pi$-antibonding orbitals of the double bond.
As the dihedral angle is rotated, the double bond is effectively broken and reformed near $90^\circ$.

\begin{figure}[htb]
\centering
\includegraphics[width=0.9\linewidth]{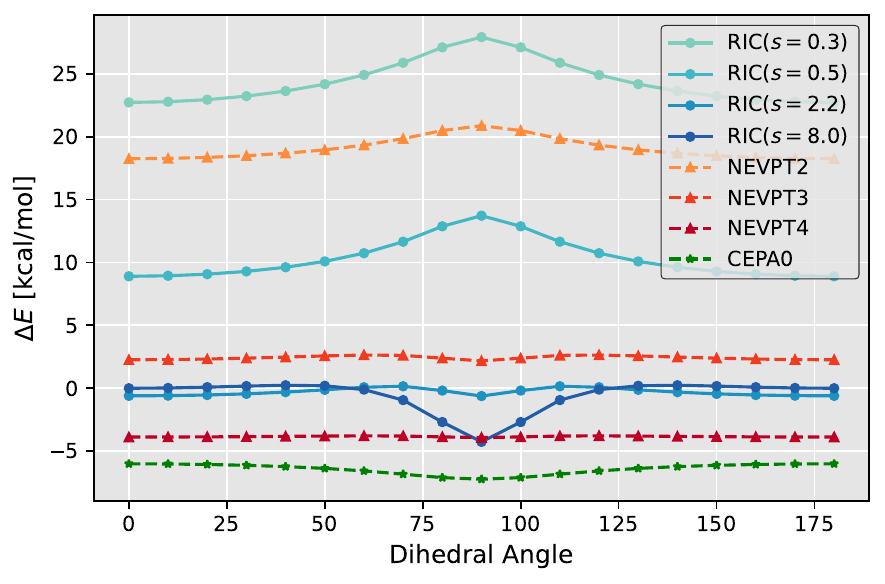}
\caption{Error during dihedral angle rotation of ethylene evaluated relative to the IC-MRCCSD method for different multireference approaches.}
\label{fig:ethe_rot}
\end{figure}

To benchmark accuracy, we assess the errors of the RIC-MRCCSD method at various flow parameter values, alongside NEVPT at second, third, and fourth order, and CEPA(0), referencing the full IC-MRCC results.
These findings are shown in Figure~\ref{fig:ethe_rot}.

A key observation is that the flow parameter significantly influences the errors in the RIC-MRCCSD method.
Recall that the parameter $s$ interpolates between the CASSCF solution at $s = 0\; E_{h}^{-2}$---lacking dynamic correlation---and the conventional many-body MRCC limit at $s \to \infty$.
Increasing $s$ therefore enhances dynamic correlation capture, but can reduce numerical stability as observed in the previous section.

The behavior of RIC-MRCCSD curves for different flow parameters mirrors somewhat the one from the NEVPT sequence.
At low perturbation order (NEVPT2 or $s < 1\; E_{h}^{-2}$), both show a pronounced concave hump around the critical angle of $90^\circ$.
As the order is increased and more dynamic correlation is recovered (NEVPT3 or $s \approx 2\; E_{h}^{-2}$), this feature is greatly reduced and, in fact, exhibits a convex profile.
NEVPT4 becomes nearly flat, indicating consistent description of the correlation along the potential energy curve with respect to the IC-MRCCSD reference energies.
In contrast, for RIC-MRCCSD, as $s$ increases, the hump becomes progressively more prevalent and eventually, at $s \geq 12 \; E_{h}^{-2}$, the method fails to converge.
We suspect that this behavior is due to the emergence of intruder states as the regularization is diminished.

Although higher-order NEVPT methods yield excellent parallelity, their absolute energies do not converge to IC-MRCC---notably, NEVPT3 is closer than NEVPT4.
Conversely, the RIC-MRCCSD scheme approaches IC-MRCCSD in absolute terms at large $s$, suggesting that the series of approximations involving the omission of expensive contractions are justified.

Evidently, the previously chosen value of $s=0.5\; E_{h}^{-2}$ in our initial study~\cite{feldmannRenormalizedInternallyContracted2024}, as well as in related DSRG approaches, proves to be not optimal for ethylene.
While $s=0.5 \; E_{h}^{-2}$ performed well for diatomic molecules, these results highlight the empirical nature and selection challenges for this parameter.
To better assess the effect of the flow parameter on the incurred error of the potential energy curve along the dihedral angle, we report the non-parallelity error
\begin{equation}
\text{NPE} = \max(\Delta E) - \min(\Delta E)
\end{equation}
which measures deviation from parallelity from the reference IC-MRCC curve.
The NPE can be considered a more relevant metric than absolute energy differences, as constant energy offsets tend to cancel once observables and properties are computed.

\begin{figure}[htb]
\centering
\includegraphics[width=0.9\linewidth]{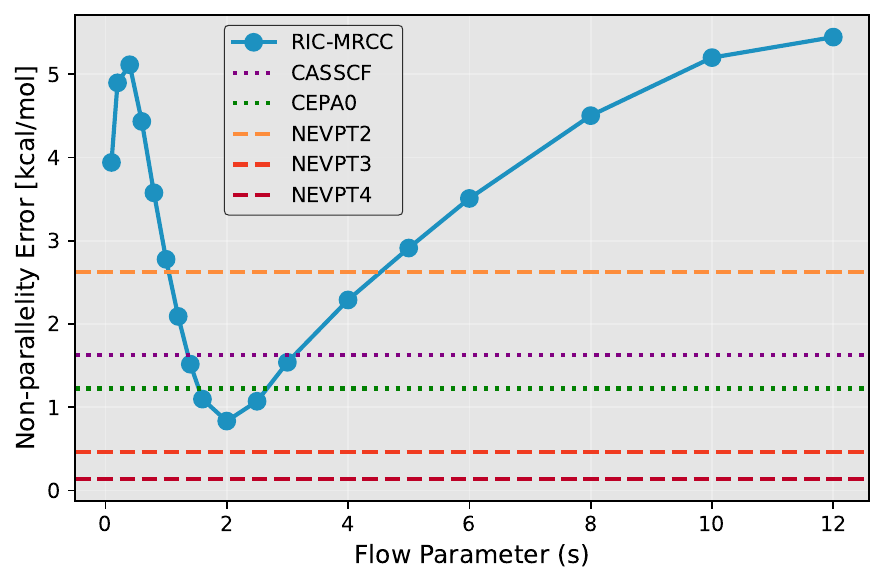}
\caption{
  Ethylene non-parallelity error along a $180^{\circ}$ dihedral rotation calculated for RIC-MRCCSD energies relative to the IC-MRCCSD curve as a function of the flow parameter $s$.
  Additionally, the CASSCF, CEPA(0) and NEVPT2, NEVPT3, and NEVPT4 NPEs are provided for comparison.
}
\label{fig:ethe_npe}
\end{figure}

Figure~\ref{fig:ethe_npe} displays the NPE across different values of $s$, alongside reference values from the other four previous dynamic correlation methods as well as the reference CASSCF value.
The RIC-MRCCSD curves attain a local maximum around the value of $s=0.5\; E_{h}^{-2}$, after which it decreases until reaching the best NPE around $s=2.0 \; E_{h}^{-2}$.
Beyond this point, the NPE increases, likely due to intruder states, and ultimately the iterations fail to converge at $s > 12 \; E_{h}^{-2}$.
The seemingly peculiar behavior at $s < 0.5 \; E_{h}^{-2}$, where the NPE seems to decrease as the flow is reduced, hence recovering less dynamic correlation, is attributable to the particularly parallel CASSCF reference, which is approached as $s \rightarrow 0 \; E_{h}^{-2}$.
Indeed, with a value of 1.63 [kcal/mol], the CASSCF solution outperforms NEVPT2 in terms of NPE (2.63 [kcal/mol]), suggesting that low-order recovery of dynamic correlation can, in fact, degrade the parallelity of the reference wavefunction.

\subsection{Size Stress Test: Vitamin B\textsubscript{12}}

\begin{figure}[htb]
\centering
\includegraphics[width=0.5\linewidth]{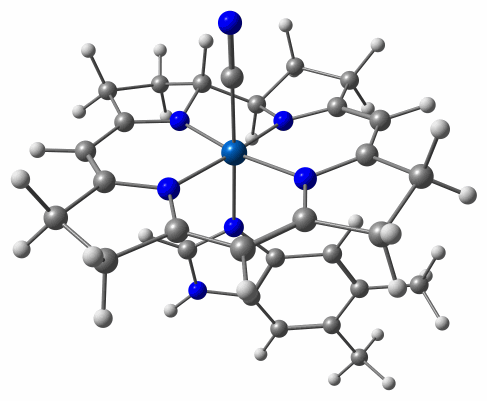}
\caption{Vitamin B\textsubscript{12} model system with its coblat transition-metal center and corrin macrocycle augmented by a histidine lower axial ligand.}
\label{fig:b12model}
\end{figure}

As a final benchmark to demonstrate the viability of the RIC-MRCCSD beyond small model systems, we report the execution time of the ground-state energy of a fairly large molecule.
In particular, we study the vitamin B\textsubscript{12} model from Ref.~\citenum{Kornobis2011Feb}, where a simplified model of molecule, containing an additional histidine lower axial ligand, was constructed from high-resolution X-ray crystallographic data.
The molecular structure of this model system is depicted in Figure~\ref{fig:b12model}.

\begin{figure}[htb]
\centering
\includegraphics[width=\linewidth]{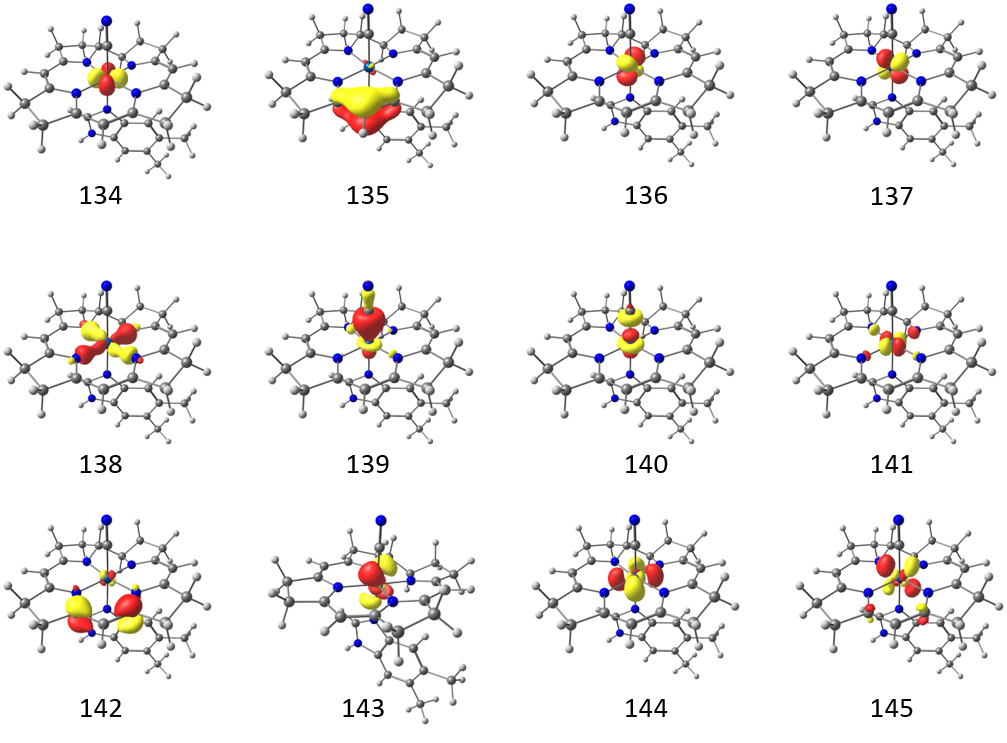}
\caption{The vitamin B\textsubscript{12} active space CAS(12, 12) comprised of the five $3d$ orbitals from the cobalt atom (134, 136, 137, 140, 141), an equatorial (138) and axial (139) bonding orbital, a $\pi$-bonding (135) and antibonding (142) pair from the corrin macrocycle and, finally, three additional $4d$ orbitals (143, 144 and 145) to account for the double-$d$ shell effect.}
\label{fig:b12cas12}
\end{figure}

For selecting the active space of the molecule, we follow the procedure from the original study~\cite{Kornobis2011Feb}, which identified an active space of 12 electrons in 12 orbitals, which are illustrated in Figure~\ref{fig:b12cas12}.

As revealed by the original study, none of the excited states of the molecule maintain the character of the corresponding CASSCF solution, following the application of the dynamic correlation scheme, suggesting that state-specific methods, such as RIC-MRCCSD, would be unreliable for describing these states.
Indeed, this state-mixing process was confirmed by us independently using the quasi-degenerate extension of the NEVPT2 method.
For the ground-state, however, the CASSCF solution comprises 85\% of the weight in the perturbed wavefunction, making it amenable state-specific schemes.
Therefore, in this section, we restrict our focus to the computation of this single state.

Our calculations applied the x2c-TZVPall~\cite{Pollak2017Aug,Franzke2019} basis set for the first coordination sphere, while all other atoms were described with the smaller x2c-SVPall~\cite{Pollak2017Aug,Franzke2019} basis set to maintain computational feasibility.
Using the frozen-core approximation, this setup yields 40 frozen, 94 internal, 12 active, and 663 virtual orbitals, representing a substantial system for high-accuracy multireference methods.
We compare the performance of the RIC-MRCCSD schemes with SR RHF-CCSD and the NEVPT2 and NEVPT4 MR methods, each run on a 32-core AMD EPYC 75F3 processor with 16 parallel MPI processes.
Runtimes and total memory usage for each approach are reported in Table~\ref{tab:b12-performance}.

\begin{table*}[htb]
\centering
\caption{Performance comparison for the vitamin B\textsubscript{12} model.}
\label{tab:b12-performance}
\begin{tabular}{l rrrr}
  \toprule
  & \multicolumn{4}{c}{Method} \\
  \cmidrule{2-5}
  & NEVPT2 & NEVPT4 & RHF-CCSD & RIC-MRCCSD \\
  \midrule
  Time & 446.3 [sec] & 7.53 [hours] & 3.49 [days] & 3.87 [days] \\
  Memory [GB] & 9.0 & 97.7 & 134.5 & 155.7 \\
  \bottomrule
\end{tabular}
\end{table*}

Although the CC methods do not match the efficiency of the NEVPT methods---particularly the highly-optimized NEVPT2 implementation in ORCA---it is encouraging that the RIC-MRCCSD method requires only marginally more time and memory than conventional RHF-CCSD.
This suggests that, with the present implementation, systems accessible to RHF-CCSD should also be accessible to RIC-MRCCSD.

\clearpage
\section{Conclusions and Outlook}\label{sec:conclusion}
In this work, we have introduced a spin-free formulation of the renormalized internally-contracted multireference coupled cluster method with single and double excitations (RIC-MRCCSD) and present its efficient implementation within the ORCA quantum chemistry package.
The implementation was accomplished by interfacing Evangelista's \texttt{Wick\&d} program---which generates the many-body residual equations in spin-orbital form---to ORCA’s native \texttt{AGE} code generator.
The resulting equations are spin adapted by \texttt{AGE} through the use of singlet-constraining relations that relate different spin sectors of the spin-orbital quantities.

We have validated fundamental properties of the method, in particular size consistency, and assessed its overall performance on a set of molecular systems including organic compounds and transition metal ions and complexes.
Our implementation showed comparable efficiency, both in terms of runtime and memory requirements, to the closed-shell single-reference coupled cluster module available in ORCA.
Moreover, since the theory involves only up to three-body cumulants, the RIC-MRCCSD approach achieves competitive performance relative to the highly optimized NEVPT2 implementation when targeting large active spaces.
As a demonstration of its applicability to extended systems, we computed the ground-state electronic energy of a vitamin B\textsubscript{12} model comprising 809 basis functions and a CAS(12,12).

With regard to accuracy, the method inherits a free parameter---the flow parameter $s$---from the closely related DSRG theory.
This parameter governs not only the accuracy but also the numerical stability of the approach.
Larger values of $s$ recover a greater portion of the dynamic correlation but may also introduce intruder states.
Such arbitrary parameters are common in multireference theories prone to intruder problems, with examples ranging from the shift parameter in CASPT2 to orthogonalization thresholds in IC-MRCC.
In DSRG, the corresponding shift parameter is generally recommended to lie within the interval $[0.1,1.0]\;E_{h}^{-2}$, with $s = 0.5\;E_{h}^{-2}$ often adopted as the default choice.
By contrast, our benchmark results indicate a significantly broader range of suitable values for the present method.
For example, to achieve convergence in transition-metal ion calculations with enlarged active spaces including the double-$d$ shell, a value of $s = 0.4\;E_{h}^{-2}$ was required to ensure convergence across all electronic states of interest.
However, for the organic molecule ethylene, this value appears too conservative, and a much larger choice of $s = 2.2\;E_{h}^{-2}$ provides improved accuracy in terms of the non-parallelity of the potential energy curves.

This work represents an initial step toward incorporating theories based on the many-body residuals into the ORCA framework, laying the foundation for the development of related approaches.
Future efforts will focus on analyzing the origin of the instabilities observed in the many-body formulation of IC-MRCC and on evaluating strategies for their mitigation.
As observed in our pilot study~\cite{feldmannRenormalizedInternallyContracted2024} and confirmed in this work, the RIC-MRCC method restricted to single and double excitations fails to consistently surpass the accuracy of established second-order multireference perturbation theories.
To address this limitation, our earlier work also proposed a perturbative triples correction, RIC-MRCCSD[T], which demonstrated strong potential in bridging this gap at moderate increase in computation cost and will, therefore, be the subject of further research.

\section*{Acknowledgments}
M.R., K.S., and R.F. gratefully acknowledge financial support by the Swiss National Science Foundation (grant no. 200021\_219616).


\providecommand{\latin}[1]{#1}
\makeatletter
\providecommand{\doi}
  {\begingroup\let\do\@makeother\dospecials
  \catcode`\{=1 \catcode`\}=2 \doi@aux}
\providecommand{\doi@aux}[1]{\endgroup\texttt{#1}}
\makeatother
\providecommand*\mcitethebibliography{\thebibliography}
\csname @ifundefined\endcsname{endmcitethebibliography}
  {\let\endmcitethebibliography\endthebibliography}{}

\end{document}